\documentclass[12pt,preprint]{aastex}

\begin{document}

\title{A Combined {\it Spitzer} and {\it Chandra} Survey of Young Stellar Objects in the Serpens Cloud Core.}

\shorttitle{A {\it Spitzer} \& {\it Chandra} Survey of Serpens}

\author{E. Winston\altaffilmark{1,2}, S. T. Megeath\altaffilmark{1,3}, S. J. Wolk\altaffilmark{1}, J. Muzerolle\altaffilmark{4}, R. Gutermuth\altaffilmark{1}, J. L. Hora\altaffilmark{1}, L.E. Allen\altaffilmark{1}, B. Spitzbart\altaffilmark{1}, P. Myers\altaffilmark{1}, G. G. Fazio\altaffilmark{1}}

\altaffiltext{1}{Harvard Smithsonian Center for Astrophysics, 60 Garden St., Cambridge MA 02138, USA.} 
\email{ewinston@cfa.harvard.edu}
\altaffiltext{2}{School of Physics, Science Centre - North, University College Dublin, Belfield, Dublin 4, Ireland.}
\altaffiltext{3}{Current address: Ritter Observatory, Dept. of Physics and Astronomy, University of Toledo, 2801 W. Bancroft Ave., Toledo, OH 43606, USA. }
\altaffiltext{4}{Steward Observatory, University of Arizona, 933 N. Cherry Ave., Tucson, AZ 85721. }

\begin{abstract}

We present {\it Spitzer} and {\it Chandra} observations of the nearby 
($\sim$260 pc) embedded stellar cluster in the Serpens Cloud Core.
We observed, using {\it Spitzer's} IRAC and MIPS instruments, in six 
wavelength bands from 3 to 70~${\mu}m$, to detect thermal emission from 
circumstellar disks and protostellar envelopes, and to classify stars using 
color-color diagrams and spectral energy distributions (SEDs). 
These data are combined with {\it Chandra} observations to examine the 
effects of circumstellar disks on stellar X-ray properties.
Young diskless stars were also identified from their increased X-ray emission. 

We have identified 138 YSOs in Serpens: 22 class 0/I, 16 
flat spectrum, 62 class II, 17 transition disk, and 21 class III stars; 
60 of which exhibit X-ray emission.
Our primary results are the following:
1.)  ten protostars detected previously in the sub-millimeter are  
detected at $\lambda < 24$~$\mu$m, seven at $\lambda < 8$~$\mu$m, 
2.) the  protostars are more closely grouped 
than more evolved YSOs (median separation : $\sim0.024$~pc), and 
3.) the luminosity and temperature of the X-ray emitting plasma around these YSOs 
does not show any significant dependence on evolutionary class.  
We combine the infrared derived values of $A_K$ and X-ray values of $N_H$ 
for 8 class III objects and find that the column density 
of hydrogen gas per mag of extinctions is less than half the standard interstellar value, 
for $A_K > 1$.  
This may be the result of grain growth through coagulation and/or the accretion 
of volatiles in the Serpens cloud core.

\end{abstract}



\section{\bf Introduction}

Surveys of molecular clouds indicate that 60\% of young stars form in  
clusters \citep{car,all2,meg2}. 
It is therefore important to study star formation in clusters  
to understand the influence of the cluster environment in the process  
of star and planet formation.  Of fundamental importance is the  
understanding of the spatial structure of clusters, and of the evolution of  
that structure.  Observations with the {\it Spitzer} and {\it Chandra} Space  
Telescopes are of great importance in the study of cluster structure  
as they provide the means to identify  young stellar objects from the  
protostellar to pre-main sequence phases, and to categorize these  
objects into the canonical evolutionary classes (class 0/I, II and  
III \citep{lad84}).  The number and distribution of sources in each of 
these classes provides unique information on the distribution of star  
formation sites, the motion of the stars after they form, and the  
dynamical state of the cluster at large.   From these studies, we can  
better understand the physical processes that govern the 
fragmentation of molecular clouds, the ensuing trajectories that the 
resulting young stars make through the cluster, and the eventual fate 
of the embedded cluster.  Clusters also serve as laboratories for the
evolution of young stellar objects, and can be used to study the  
evolution of disks around young stars and the evolution
of hot X-ray emitting plasma commonly found around such stars 
(see for example: \citet{gut,hai,pre3,her})

In this paper, we describe a detailed study of the population  of young 
stellar objects in the Serpens cluster identified by observations with {\it Spitzer}  
and {\it Chandra}.  The Serpens region is an example of a very young, deeply 
embedded cluster, containing a number of protostars \citep{har,hur,eir1,eir2,kaa,tes2}. 
The embedded cluster is heavily extinguished, with a peak extinction exceeding 40 
magnitudes in the visual.
At a distance of $\sim$260~pc, the Serpens cloud core is one of the nearest regions 
of clustered star-formation to the Sun (see \citet{str} for a discussion of the 
distance to Serpens). This makes it an excellent 
candidate for study with {\it Spitzer} as it is close enough to both resolve the individual 
members and to detect the lowest mass members to below the hydrogen-burning 
limit. 
Furthermore, the Serpens cluster is rich in protostars.
Sub-millimetre and millimetre observations of the region, at 
450~${\mu}m$, 850~${\mu}m$ and 3~mm, identify at least 14 objects in the 
Serpens Cloud Core \citep{tes1,tes2,dav1,hog,har}.
In this paper, we extend the sample of known young stellar objects in Serpens using 
the high sensitivity of {\it Spitzer} and {\it Chandra}.    We  then discuss three topics: 
the mid-IR spectral energy distributions of protostellar objects detected in the 
sub-millimeter, the spatial distribution of the Serpens cluster members, and 
finally, the X-ray properties of the young stellar objects as a function of their evolutionary class.

\section{\bf IRAC \& MIPS Data Reduction}

\begin{figure}
\plotone{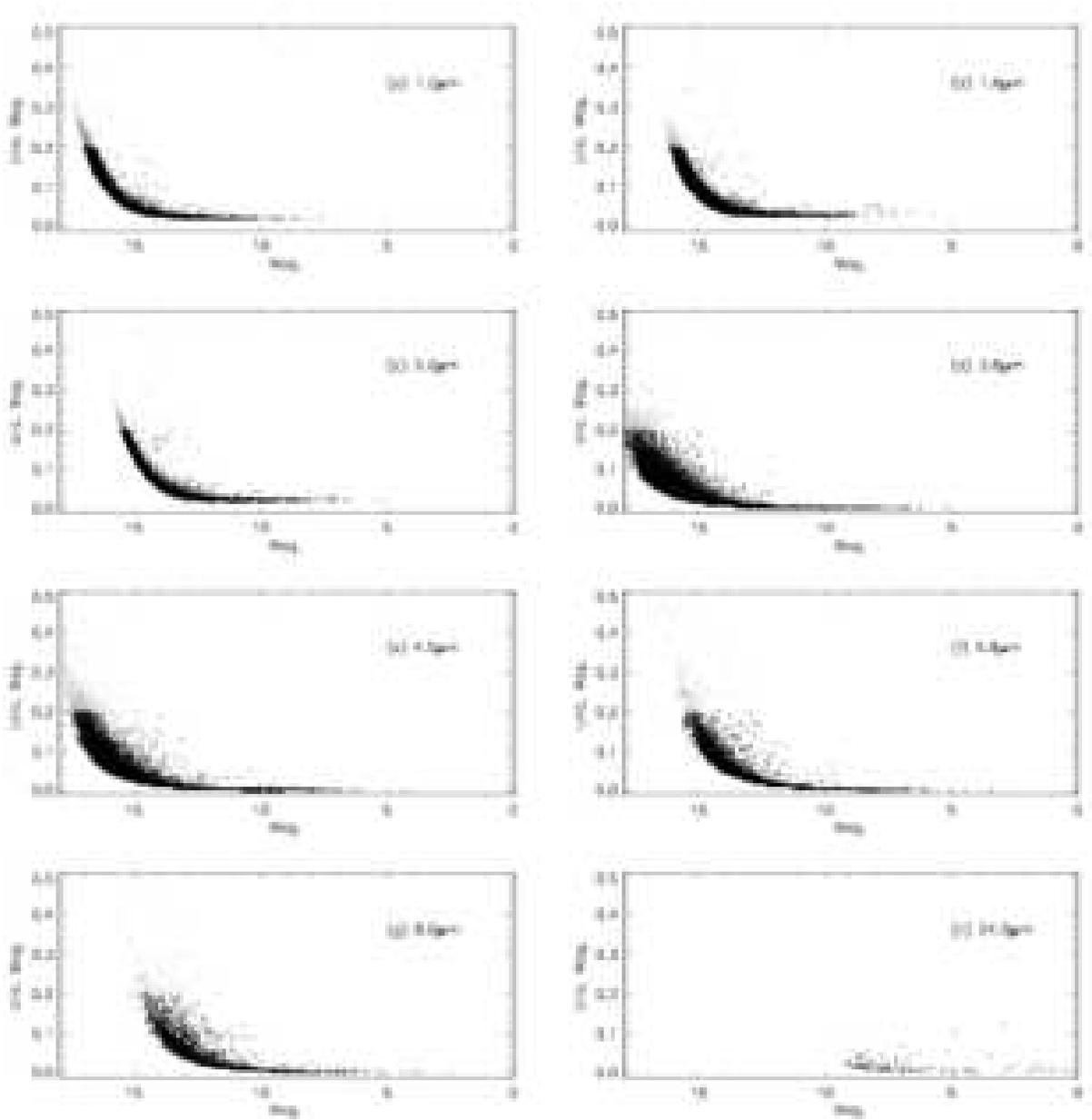}
\caption{ The magnitude vs magnitude uncertainty for each of the eight IR bands 
 used in the study. The black points indicate those sources having 
 errors in magnitude less than 0.2 mag, the grey points those with larger errors. 
 This cutoff was applied to all selection techniques.
 (a) $J$, (b) $H$, (c) $K_s$, (d) 3.6~${\mu}m$, (e) 4.5~${\mu}m$, (f) 5.8~${\mu}m$, 
 (g) 8.0~${\mu}m$, (h) 24~${\mu}m$.   The near-IR data is from the 2MASS catalogue.  }
\protect\label{fig3}
\end{figure}

We have obtained {\it Spitzer} images of the Serpens Cloud Core in six
wavelength bands: the 3.6, 4.5, 5.8 and 8.0~$\mu$m bands of the
InfraRed Array Camera (IRAC; \citet{faz}) and the 24~$\mu$m and
70~$\mu$m bands of the Multi-band Imaging Photometer for {\it Spitzer} (MIPS; 
\citet{rie}).  The photometry extracted from these data was
supplemented by $J$, $H$ and $K_S$-band photometry from the 2MASS point
source catalog \citep{skr}, resulting in data in nine photometric
bands spanning 1-70~$\mu$m.  The uncertainties for the point sources as
a function of magnitude are shown in Fig.~\ref{fig3}.  In the
following analysis, only data with uncertainties $\le 0.2$~magnitudes were
used. Below, we describe the observations, image
reduction and photometry for the {\it Spitzer} data.

\subsection{IRAC:}

\begin{figure}
\plotone{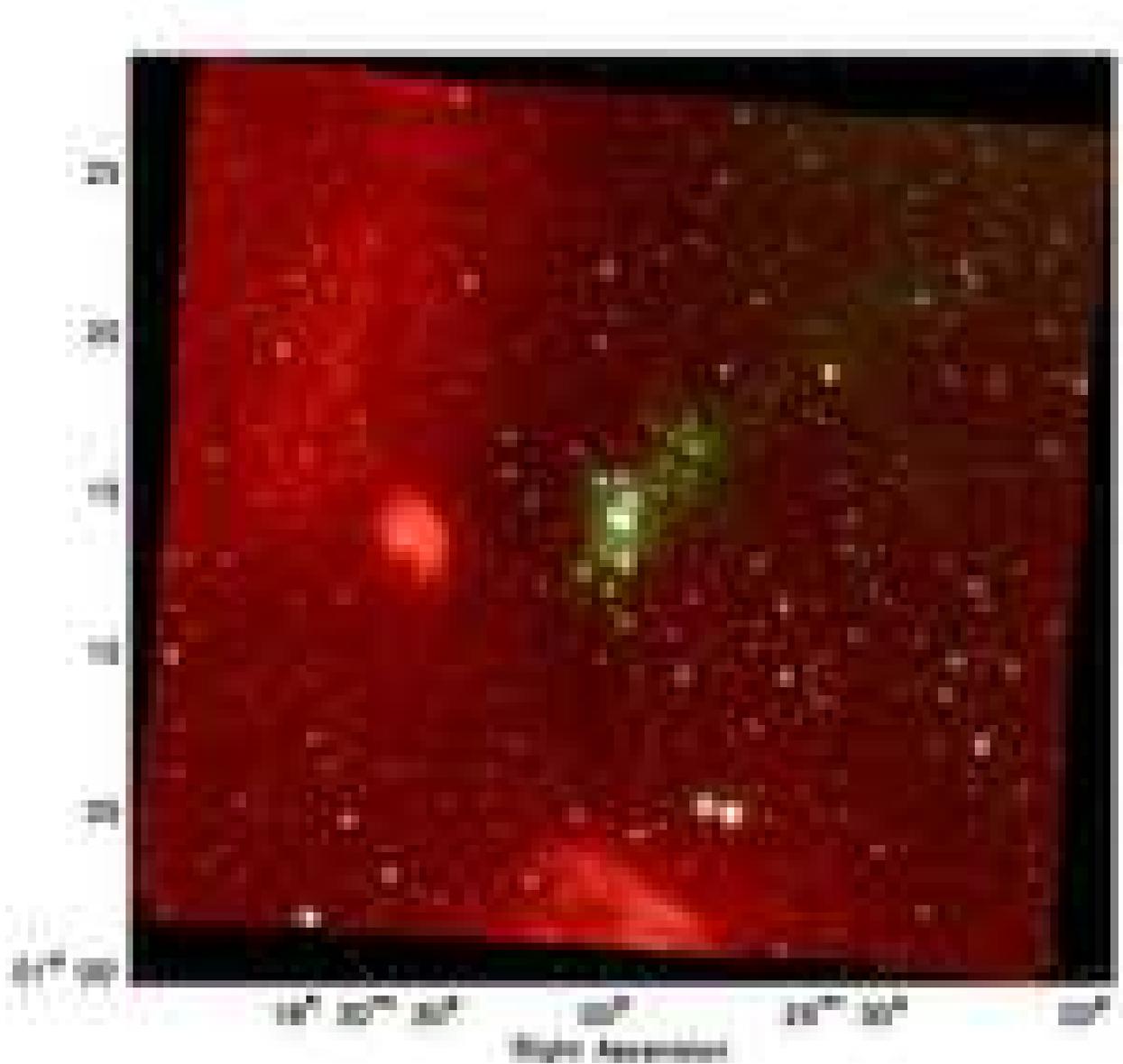}
\caption{  The Serpens cloud core region. A three band false color image of Serpens 
 using IRAC on {\it Spitzer}. Blue is 3.6~${\mu}m$, Green is 4.5~${\mu}m$, and Red is 8.0~${\mu}m$.
 Much of the reddish hue in the image is due to the diffuse PAH emission. 
 Emission from shocked hydrogen is visible in green.  The light blue-green is scattered light.  }
\protect\label{fig1}
\end{figure}

The {\it Spitzer} Space Telescope has observed the Serpens cluster  
with the IRAC instrument at two different epochs. The first epoch was
on 1st April 2004; these observations were part of the Guaranteed Time
Observation program PID 6, {\it The Structure and Incidence of Embedded 
Clusters}.  The second epoch was on 5th April 2004, during which {\it 
Spitzer} executed observations from PID 174 from the {\it cores to disks}  (c2d)
legacy program.  The mosaicked regions cover a $29' \times 33'$ field
in all four of the IRAC wavelength bands (Fig.\ref{fig1}).  The field
is centered on the main Serpens cluster in the Northern Cloud Core.
The observations were taken in the high dynamic range mode with a 12
second frame time. In each epoch, two dithered 10.4 second frames and
two dithered 0.4 second frames were obtained at each map position.

The analysis used the basic calibrated data (BCD) FITS images from the
S11.4 pipeline of the {\it Spitzer} Science Center.  These data were
combined into mosaics using the MOPEX program.  The 3.6~$\mu$m and 
4.5~$\mu$m images were pre-processed by 
using the IDL-based pulldown corrector by Lexi Moustakas \footnote{http://ssc.spitzer.caltech.edu/irac/pulldown/}, 
and then additional interactive processing steps
were applied to locate and remove the remaining column pulldown and mux
bleed artifacts in the 3.6~$\mu$m and 4.5~$\mu$m images, and banding effects in the 
5.8~$\mu$m and 8.0~$\mu$m images \citep{hor}.  The overlap correction module was used to minimize
instrumental offsets between adjacent frames, and the mosaics were
constructed on a pixel scale of 0.8627 $''$/pixel, or $1/\sqrt{2}$ the IRAC
pixel size. The typical total integration time is 41.6 seconds per
pixel, although in some regions near the edges of the mosaic, the
integration time is 20.8 seconds.
The calibration uncertainty across the IRAC bands is estimated at $0.02$~mag 
\citep{rea}.
In addition, there are $\sim$5\% position dependent variations in the calibration 
of point sources in the flat-fielded BCD data; these have not been corrected for 
in our data.  Furthermore, these systematic uncertainties have not been added 
into the random photometric uncertainties reported in this paper.

We used a custom IDL program for point source finding; this program
was a heavily modified version of the DAOfind program in the IDLPHOT
package \citep{lan}. This program first creates a smoothed version of the
mosaic by convolving with a Gaussian with  $\sigma =  4$~pixels, 
approximately double the FWHM of the point sources, 
and then subtracts the smoothed version from 
the mosaic to filter out the extended nebulosity. 
The noise in a $8 \times 8$ pixel region around the source was calculated 
to assess the combined instrumental noise, shot
noise and noise from the spatially varying extended nebulosity. 
Point sources with peak values more than 5 $\sigma$ above the background 
were considered candidate detections.  
After the point sources were identified,
aperture photometry then was obtained using the aper.pro program in
the IDLphot package. An aperture radius of 2.8 pixels ($2.4''$) was
used, and a sky annulus from 4.2 to 8.4 pixels ($3.6''$ to $7.2''$)
was used to measure the contribution from extended emission in the
aperture.  
The zero points for these apertures were, 17.8204, 17.3025, 16.7408, 
and 15.9440~mag for the 3.6, 4.5, 5.8, and 8.0~$\mu$m bands, respectively, 
in the native BCD image units of MJy pixels sr$^{-1}$.

There were 25853, 22703, 6595, and 3946 detections in the 3.6, 4.5, 5.8, and 8.0~${\mu}m$
respectively, with uncertainties less than 0.2~mag,  with 2758 of these detected in all four 
IRAC bands. The sensitivity of the 8.0~${\mu}m$ 
band is the limiting factor in four band detections: it suffers from lower 
photospheric fluxes, higher background emission, and the presence of bright, 
structured nebulosity in some parts of the image. The detection threshold in 
the 3.6~${\mu}m$ band was 17 magnitudes - far below the hydrogen burning limit 
for a 1~Myr old star  at the distance of Serpens ($\sim$12~mag at 3.6~$\mu$m \citep{bar}).

\subsection{MIPS:}

{\it Spitzer} observed the Serpens cluster twice with the Multiband
Imaging Photometer for {\it Spitzer} (MIPS; Rieke et al. 2004). 
In both epochs, the medium scan rate was used.  For the first epoch, part of 
the Guaranteed Time Observation program PID 6, taken on the 6th April 
2004, six 0.5 degree scan legs with full-array cross-scan offsets were used. For 
the second epoch observations (PID 174), 12th April 2005,  12 scan 
legs of length 0.5 degrees and half-array cross-scan offsets were used.
In both epochs, the total map size was approximately 0.5 x 1.5 degrees including 
the overscan region. 
All three MIPS bands are taken simultaneously in this mode; because the data 
at 160 micron are saturated, we do not consider it in our subsequent analysis.  
The 2nd epoch map had full 70~$\mu$m coverage, while the first epoch
mapped only six $150''$ wide bands, covering half of the map.  
Both epochs were combined to form the final maps.  
The typical effective exposure time per pixel is about
80 seconds at 24~$\mu$m and 40 seconds at 70~$\mu$m. The MIPS images
were processed using the MIPS instrument team Data Analysis Tool,
which calibrates the data and applies a distortion correction to each
individual exposure before combining into a final mosaic \citep{gor}. 
The data were further processed using various median
filters to remove saturation effects and column-dependent background
structure.  The resulting mosaics have a pixel size of $1.25''$ at 24~$\mu$m 
and $4.95''$ at 70~$\mu$m.

Stars in the 24~$\mu$m mosaic were then identified using PhotVis \citep{gut2}, 
which searched for all point sources with peaks 5 times the RMS noise.
Aperture photometry in a 5 pixel aperture was performed on these
sources with PhotVis and the photometry was corrected to a 12 pixel aperture
radius, with a sky annulus from 12-15 pixels. 
Adopting a conversion of 6.711 Jy per $DN s^{-1} pix^{-1}$, multiplying
by $1/4$ to correct for the smaller mosaic pixels and by 1.146 to correct
from a 12 mosaic pixel aperture to an infinite aperture, and using a zero
magnitude flux of 7.3 Jy, the aperture photometry was converted to
magnitudes.  Due to the larger $5''$ FWHM of the MIPS 24~$\mu$m Point
Spread Function (PSF), we performed PSF fitting photometry on all the
detected sources using the IDL version of DAOPHOT in the IDLPHOT
package.  The PSF was generated from six bright (3-6 magnitude) stars
in the image; these stars were chosen to be bright and uncontaminated
by nebulosity. 
PSF fitting photometry was then performed on the point source detections 
using the nstar.pro routine in IDLPHOT. The nstar routine requires the input 
of aperture photometry. Sky values were taken from an annulus from 12 to 15 
pixels. These photometry are then adjusted by fitting the PSF to the image data 
and scaling accordingly. A total of 269 sources were extracted 
with uncertainties less than 0.2~mag.
Three of the sources, numbers 11, 35, and 36, were saturated at 24~$\mu$m.  
The fits of these data were adjusted by scaling the point spread function 
manually, subtracing the PSF from the image,  and visually inspecting the 
residual in the wings of the PSF.  The tabulated magnitudes produced the 
lowest apparent residual. Since raising/lowering the magnitudes by 0.1~mag 
would produce a distinct over/undersubtraction in the residual image, we 
adopted an uncertainty of 0.1 magnitude for these values.

The 70~$\mu$m photometry was extracted from the mosaic using a similar 
procedure to the 24~$\mu$m photometry. Aperture photometry was performed on the 
sources in a 3.66 pixel aperature, with a sky annulus from 3.66-7.92 pixels. 
The conversion used was 0.675 Jy per $DN s^{-1} pix^{-1}$, with a factor of 1.927 
correcting from a 3.66 pixel to an infinite aperture, and a zero magnitude flux 
of 0.775 Jy.  PSF photometry was not performed on the point sources as there 
were not enough detections uncontaminated by nebulosity to generate the PSF.
The FWHM of the PSF at 70~$\mu$m is 18$''$. The typical uncertainty on the 
70~$\mu$m flux is 15\%.

\section{{\it Chandra} X-ray Data Reduction}
The X-ray data were taken from the $Chandra$ ANCHORS (AN archive of CHandra 
Observations of Regions of Star formation\footnote{http://cxc.harvard.edu/ANCHORS/}) 
archive, obsid 4479.
The raw X-ray data are dominated by events of non-astrophysical origin.
To remove these, the raw data were processed using the ${\it Chandra}$ X-ray 
Center's standard processing version 6.13.2 (July 2005).
Using acis\_process\_events on the level 1 events file,
a gain correction (conversion from pulse heights to X-ray energy) is applied 
from CALDB 2.21.
The CTI correction for pulse heights distorted by Charge transfer inefficiency
and VFAINT background cleaning to remove soft cosmic rays were also applied.
Next an energy filter was applied to remove photons above 8~keV and below 300~eV 
which are typically not of stellar origin.
Finally events with bad grades and bad status (grades 1, 5 and 7,
status $>$0 indicative of X-ray signals) and bad time intervals were
filtered using the CIAO tool {\it dmcopy}.
Time-dependent gain corrections were applied and acis\_process\_events rerun.
While there were over 3$\times10^6$ level 1 events, our ``cleaned'' data file
used for analysis contained 171,973 events.

Due to vignetting and small gaps between the chips of the I array
the effective collecting area on each part of the sky differs.  The effective
area also depends on the energy of the photon.  The exposure map
corrects for the changes in the effective area. 
An exposure map was created using {\it merge\_all}, accurately representing the
average effective area for a 1.7~keV photon. We chose this single energy for
the exposure map since it is intermediate between the maximum of the effective
area of the HRMA/ACIS system and an estimated mean source energy of $\sim$ 2.0~keV.  
The exposure map is later applied automatically by CIAO tools extracting count rates 
and spectra.

\subsection{Source Detection}
To perform source detection, the data were split into two event lists, one
to concentrate on cooler sources with limited noise, consisting of
photons with energies between 0.5 and 2.0~keV.  The other list
contained photons of higher energy between 2.0 and 7.5~keV.
WAVDETECT was used for source detection on our cleaned  and separated events 
list to
identify sources across the entire I array. Threshold significance was set to
detect sources down to 3.5 sigma and the data were
searched on scales of 0.5 to 16$''$.  With these settings, a false detection
rate of $<1\%$ is expected. The source detection lists were combined
to make a single source list of 88 Sources.
IN comparison, \citet{gio} construct their source list from the same data 
using different filtering criteria 
on the bottom 10\% of the data. There is $\sim$95\% agreement between the two 
source catalogues.

 From the location of the 88 sources, regions were calculated which
would contain 95\% of the X-ray energy of each sources.
The regions are based on the ${\it Chandra}$ PSF and chip position for
each source.  The routine {\it mk\_psf} was used to obtain images of the PSF at
various off-axis angles $\theta$~(arcmin), and
rotation angles $\phi$~(degrees), around the ACIS array.  At each source
location an ellipse was generated to enclose 95\% of the total X-ray energy
following Wolk et al. (2006).

\subsection{Spectral Analysis}

       Analysis of the X-ray spectrum of each source was performed to determine
the bulk temperature of the corona and the intervening column of
hydrogen.  For each source with over 25 counts,
Source and background pulse height distributions in the total band
(0.3-8.0 keV) were constructed.
 The final fits were done with CIAO version 3.1.0.1 using
   the CIAO script {\it psextract} to extract source spectra and to
create an Ancillary Response Function (ARF) and Redistribution Matrix Function
(RMF) files which correct for the detector response at each location.
Model fitting of spectra was performed using {\it Sherpa}  
\citep{fre}. 
The data for each source were grouped
into energy bins which required a minimum of 8 counts per bin and background
subtracted. The optimization method was set to Levenberg-Marquardt and
$\chi$--Gehrels statistics were employed.
An absorbed one--temperature ``Raymond--Smith'' plasma \citep{ray},
was fitted using a two step fitting procedure. Initial conditions were set so that
$nH = 1.0 \times 10^{21} cm^{-2}$ and $ kT = 1.0$ keV. Then an initial fit was made
with an absorbed thermal blackbody model.
These fit results were then used as initial conditions for the
absorbed Raymond--Smith plasma model.

\section{Band Merging}

To generate a catalog of the photometry from all detected sources in
the Serpens field, the photometry from the combined IRAC, MIPS and
{\it Chandra} data were merged.  In addition, the data were also merged with
the 2MASS point source catalog to provide $J$, $H$ and $K_S$-band
photometry for each detected {\it Spitzer} source \citep{skr}.  Sources
observed in different wavelength bands which were located within $1''$
of each other were considered to be the same source; if while
comparing two bands, multiple sources in one of the bands satisfied
this criteria, the closest source was chosen.

\begin{figure}
\epsscale{.80}
\plotone{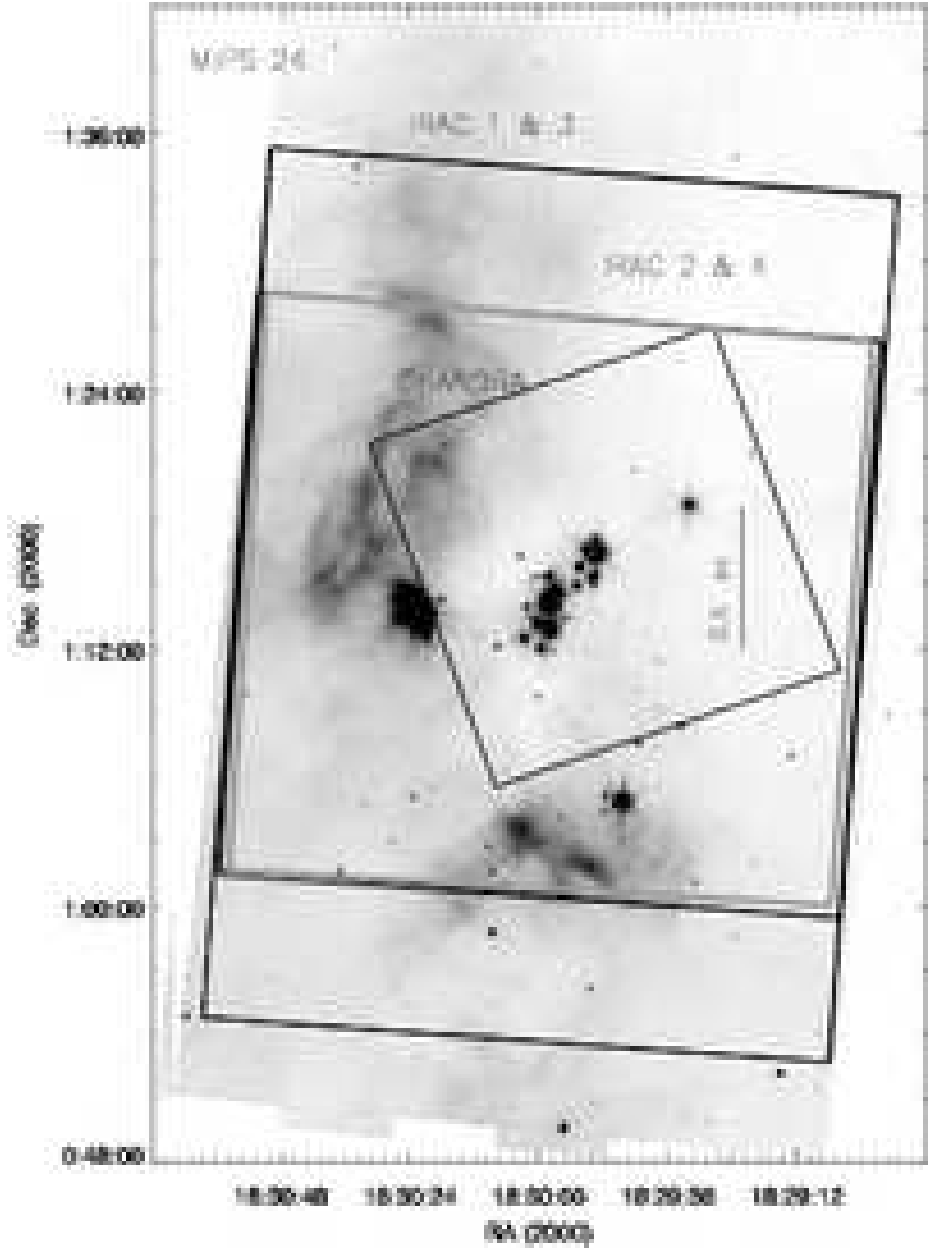}
\caption{   A grey-scale image of the Serpens cloud core at 24~${\mu}m$ showing the
 entire field of view. Overlaid, in black, are the outlines of the fields 
 of view of the IRAC 3.6 \& 5.8~${\mu}m$ (Chs. 1 \& 3) and 4.5 \& 8.0~${\mu}m$ (Chs. 2 \& 4), 
 and the {\it Chandra} field of view (the IRX-field). The IR-field, which contains complete 
 coverage in all the infrared bands, is outlined in gray.     }
\protect\label{fig2}
\end{figure}

The field of view (FOV) of each instrument did not cover the same
region of the cluster, being most limited by the IRAC FOV in the
infrared, and overall by the {\it Chandra} FOV.  The MIPS data covered the
entire IRAC field, while the 2MASS survey is not spatially limited.
Furthermore, the IRAC detectors are split into two groups, channels 1 \&
3 (3.6 \& 5.8~${\mu}m$) and channels 2 \& 4 (4.5 \& 8.0~${\mu}m$), whose FOVs are 
offset from one another by $6.5'$.  The field studied in the remainder of this paper is the
overlap region of all five {\it Spitzer} bands, the IR-field (Fig.\ref{fig2}). 
The size of the overlap regions is 26$'$ $\times$ 28$'$.   In comparison, the {\it Chandra}
image of Serpens covered a 17$'$ $\times$ 17$'$ field of view  (the IRX-field), 
centered on the northwestern edge of the cluster,  $\sim$2$'$ from the center of the IRAC FOV (Fig.~\ref{fig2}). 
X-ray data is not available for all the sources in our catalog.

\section{\bf Identification of Young Stellar Objects}

The Serpens field contains 19,181 sources with a detection in at least one band with 
photometric uncertainty $< 0.2$ in the IR-field (122 having detections in all five bands); 
however, only a small fraction of these are bona fide YSOs. Three methods were applied to 
these data to identify possible young stellar objects: selection of stars with IR excesses 
on IR color-color diagrams, identification of X-ray luminous YSOs by comparison of X-ray 
sources with IR detections, and finally, a search for extremely red mid-IR sources among 
the detections.

\subsection{Detection by InfraRed Excess}

\begin{figure}
\epsscale{1.0}
\plotone{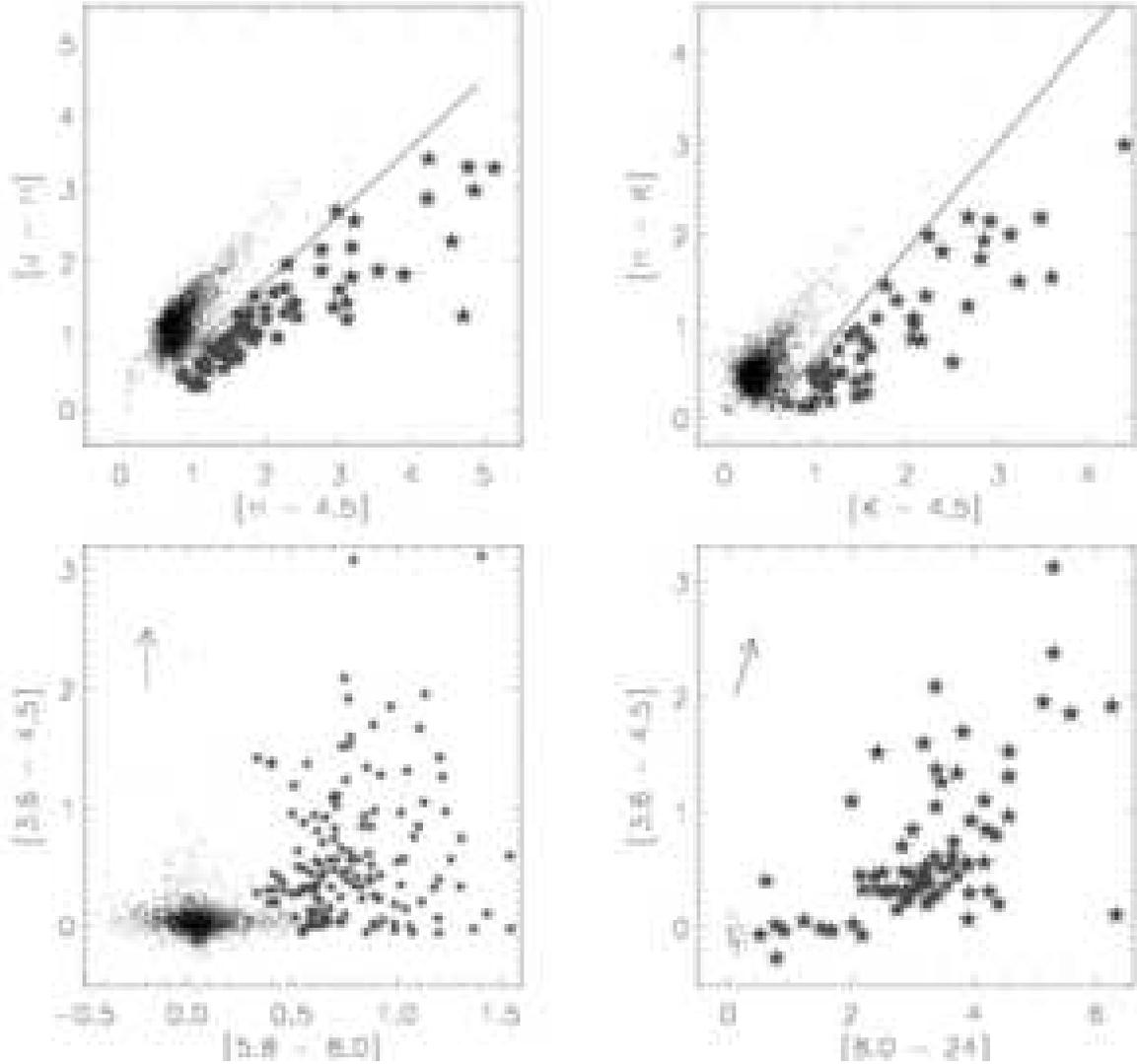}
\caption{ The four color-color diagrams used to identify young stellar
objects.  {\it Top Left}: the $J-H$ vs. $H-[4.5]$ diagram. The solid
line shows the reddening vector for these wavelengths, objects with
$H-[4.5]$ greater than one sigma below the reddening vector are
considered to have an excess. The black dots indicate those objects
classified as having no excess in this diagram, the grey stars indicate
those sources with an excess.  {\it Top Right}: The $H-K$ vs. $K-[4.5]$
diagram. Similar to above, with stars indicating those sources with
excess in their $K-[4.5]$ colors.  {\it Bottom Left}: The IRAC four
color-color diagram, [3.6] - [4.5] vs. [5.8] - [8.0]. The locus for
field stars lies on the origin, the spread in the [5.8] - [8.0]
colors is in part due contamination from nebulosity and from
contamination from star-forming galaxies with strong PAH emission in
the 5.8 and 8.0~$\mu$m bands.   {\it Bottom Right}: 
A MIPS and IRAC color-color diagram: [3.6] - [4.5] vs. [8.0] -[24.0]. 
Again, the field star locus is at the origin. 
This diagram is particularly efficient at identifying
`transition' disk objects, those having little excess shortward of
8.0~${\mu}m$.  A reddening vector of $A_K=5$ is shown in both the 
IRAC and the IRAC-MIPS diagrams. }
\label{fig7}
\end{figure}

Young stellar objects can be identified by their excess emission at IR
wavelengths. This emission arises from reprocessed stellar radiation
in the dusty material of their natal envelopes or circumstellar disks.
The infrared identification of YSOs is carried out by identifying
sources that possess colors indicative of IR excess and distinguishing
them from reddened and/or cool stars \citep{meg,all,gut1,muz}.  
The main limitation of this method is the
contamination from extragalactic sources such as PAH-rich star-forming
galaxies and AGN; both of these have colors similar to young stellar
objects. Four color-color diagrams were used to determine the color
excess of the sources: an IRAC  [3.6] - [4.5] vs. [5.8] - [8.0]
diagram, an IRAC-MIPS [3.6] - [4.5] vs. [8.0] - [24] diagram, and two
IRAC-2MASS diagrams J - H vs. H - [4.5] and H - K vs. K - [4.5], 
Fig.\ref{fig7}.  In the following analysis we required all photometry
to have uncertainties $< 0.2$~mags in all bands used for a {\it
particular} color-color diagram.  The numbers of sources for each
diagram which satisfy this criteria are 2417, 122, 3748, and 4251,
respectively.

\subsubsection{IRAC Color-Color Diagram}

To identify sources with true IR excesses, it is necessary to
distinguish between reddened or cool stars and those with excess
emission arising from heated dust grains.  A reddening law in the IRAC bands was
determined by Flaherty et al.(2006), which shows the [5.8] - [8.0]
color to be particularly insensitive to reddening and stellar
temperature, and is thus a reliable measure of excess emission due to
dust.  Sources with a color $> 1 \sigma$ beyond $[5.8] - [8.0] > 0.2$
mag are likely to possess excesses and not to be reddened or cool
stars (Fig.\ref{fig4}).

\begin{figure}
\plotone{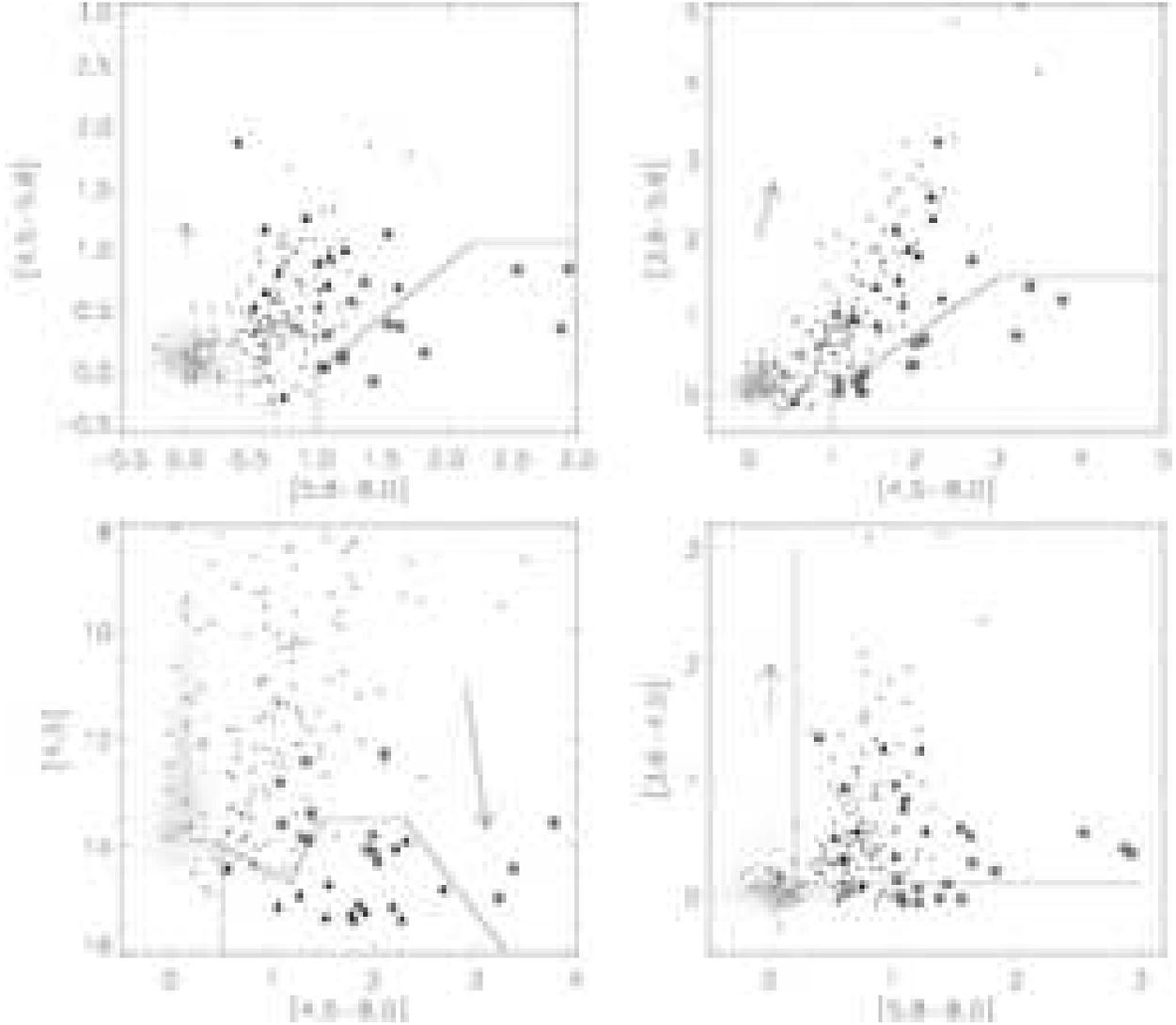}
\caption{ The above diagrams show the color and magnitude criteria
 used to remove likely star-forming galaxy and AGN contamination \citep{gut07}.  
 The grey dots are all sources with uncertainties $<
 0.2$~mag in the relevant bands, the small squares are the YSOs, the
 large grey squares are sources selected as PAH-rich star-forming
 galaxies, and the black triangles are sources selected as AGN.  The
 [4.5] - [8.0] vs. [5.8] - [8.0] ({\it above left}), [3.6] - [5.8] vs. [4.5] -
 [8.0] ({\it above right}), and [3.6] - [4.5] vs [5.8] - [8.0] ({\it bottom left}) 
 diagrams are used to find the star-forming galaxies.  In these cases, the
 lines show the criteria used to identify the star-forming
 galaxies. The AGN show colors much more similar to the YSOs and are
 instead identified by their faintness.  The magnitude criteria used
 to find AGN is shown as a line in the [4.5] vs. [4.5 - [8.0]
 color-magnitude diagram. 
The IRAC color-color diagram ({\it bottom right}) is used to select out pure photospheres and 
stars contaminated by nebulosity.
Excess sources have colors $> 1 \sigma$ beyond 
 $[3.6] - [4.5] > 0.1$ and $[5.8] - [8.0] > 0.2$. A 5 $A_K$ reddening 
 vector, from Flaherty et al.(2006), is overlaid on each plot. }
\protect\label{fig5}
\end{figure}

Extragalactic sources such as PAH rich star-forming galaxies and AGN
will also satisfy this criteria \citep{ste}.
Sources with a color more than $1~\sigma$ below $[3.6] - [4.5] < 0.1$ were considered 
galaxies and were removed; those with a 24~${\mu}m$ detection are reconsidered separately.
Gutermuth et al.(submitted)  have recently developed a method for substantially 
reducing extragalactic contamination built on the Bootes Shallow Survey data \citep{eis} 
and the {\it Spitzer cores to disks} legacy program methods \citep{jor,har} .   
The galaxies are eliminated either by their colors, which unlike YSOs are 
often dominated by PAH features in the 5.8 and 8.0~${\mu}m$ bands, or 
by their faintness. 
AGN are typically much fainter than the YSOs found in nearby star-forming
regions and are removed by this criterion. It should be noted that very faint 
or embedded flat spectrum sources may be erroneously filtered by this 
method also.  
Fig.\ref{fig5} shows the color-color and color-magnitude
diagrams used to identify star-forming galaxies with strong PAH
emission and AGN.

Nebulosity may also result in the misidentification of sources.
There are 55 sources with excesses in the [5.8]-[8.0] color, but
little evidence of an excess in the [3.6]-[4.5] color.  Examination of
the spectral energy distributions for these sources (Sec.~6) showed
that these sources exhibit excess emission only in the 8.0~${\mu}m$ band.
Subsequent analysis showed that seventeen of these also showed excesses in
the 24~${\mu}m$ band, but 38 showed no 24~${\mu}m$ detection.  The seventeen
24~${\mu}m$ excess sources followed the distribution of the other bona
fide YSOs in the field, while the 38 sources with weak 8.0~${\mu}m$ and
no 24~${\mu}m$ excess were found in regions of bright 8.0~${\mu}m$ nebulosity to
the East of the cluster (Fig.~2).  These sources appear to have their
8.0~${\mu}m$ photometry contaminated by the nebulosity and were removed
from the sample.  In general, all sources with [5.8]-[8.0] excesses
which do not show an excess at 24~${\mu}m$ excess or a $[3.6]-[4.5] >
0.1$ are removed.

In total, 146 potential YSOs were identified in the IRAC color-color
diagram.  Of these, 62 are likely to be contaminants and have been
removed from the source list.

\subsubsection{IRAC-2MASS Color-Color Diagrams}

The shorter wavelength 3.6 and 4.5~${\mu}m$ IRAC bands are much more
sensitive to stellar photospheres than the longer wavelength 5.8 and
8.0~${\mu}m$ bands; where the shorter wavelength bands have 17,385 and
15,539 detections  with $\sigma \le 0.2$, respectively, the longer wavelength bands having
4,670 and 2721 detections. Hence many sources cannot be
characterized with the IRAC color-color diagram.  For this reason, it
is important to develop methods to identify sources with infrared
excesses that rely only on the shorter wavelength bands.  Since we
cannot distinguish between reddening and infrared excess from
circumstellar dust with only two bands, we combine the near-IR data
from the 2MASS point source catalog with the 3.6 and 4.5~${\mu}m$ band photometry 
\citep{skr}.  This provides the ability to detect objects which
are too faint for detection in the 5.8 and 8.0~${\mu}m$ bands.  In
particular, we concentrate on the $J-H$ vs. $H-[4.5]$ diagram and the
$H-K$ vs $H-[4.5]$ diagram.  These diagrams take advantage of the
excellent sensitivity of {\it Spitzer} in the 4.5~$\mu$m band and the
stronger infrared excess emission at 4.5~$\mu$m compared to that at
shorter wavelength \citep{gut2}.   For highly reddened sources which are not
detected in the $J$-band, the $H-K$ vs. $K-[4.5]$ diagram can be used.
It should be noted that the IRAC 3.6 and 4.5~${\mu}m$ data can detect
sources too faint or reddened to have been detected by 2MASS.  Deeper
near-IR imaging is needed to detect these sources in the $J$, $H$ and
$K_S$-bands.

\begin{figure}
\plotone{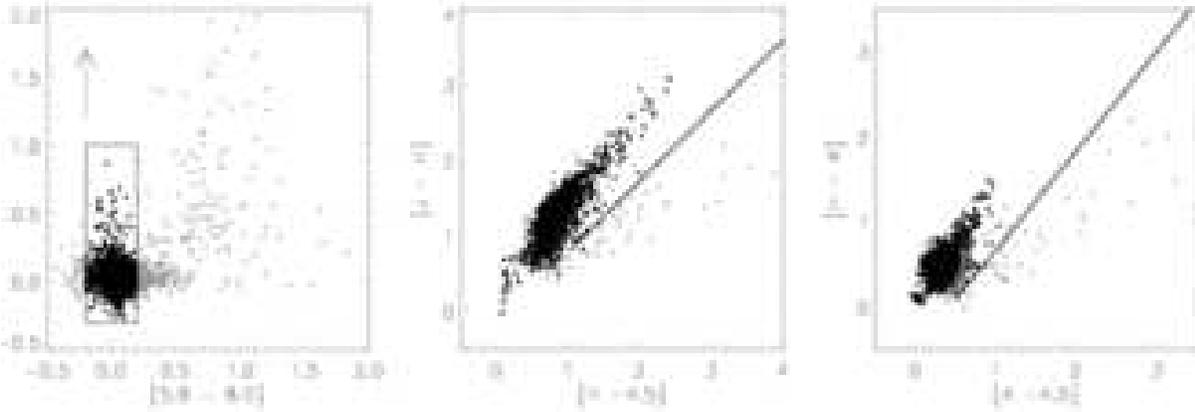}
\caption{These figures illustrate the method used to establish the
locus of reddened stellar photospheres on the IRAC-2MASS color-color
diagrams.  Only data with uncertainties 0.1~mag or less in all IRAC and
2MASS bands were plotted.  The displayed reddening law was derived
from this data by \citet{fla}.  In each diagram, the black
dots are the selected reddened photospheres, while the grey dots are
the sources which show infrared excesses or large photometric scatter.
Left: The IRAC [3.6] - [4.5] vs. [5.8] - [8.0] color-color diagram was
used to select stars which do not show infrared excesses in the IRAC
bands. The square indicates the location of the field star
locus on this color-color plot.  The reddening vector is for an $A_K
= 5$.  Middle: The black dots show the locus of reddened stellar
photospheres plotted on the $J - H$ vs. $H - [4.5]$ color-color diagram.
The reddening vector serves as our adopted dividing line between
reddened photospheres and stars with infrared excesses; sources more
than 1~$\sigma$ to the right of this vector are identified as having
an infrared excess.  Three sources out of the 1625 are misidentified 
as excess as their IRAC [3.6]-[4.5] vs [5.8]-[8.0] colors show them 
to be pure photospheres without infrared excess.
Right: The distribution of reddened photospheres
on the $H - K$ vs. $K - [4.5]$ color-color diagrams.  Again, the
displayed reddening vector is our adopted dividing line between pure
photospheres and stars with infrared excesses. One source out of 1625 is 
misidentified as an excess as its IRAC colors are indicative of a pure 
photosphere.  }
\protect\label{fig4}
\end{figure}

We calibrated the combined IRAC-2MASS color-color diagrams using
a sample of stars which show no evidence for infrared excesses out to
8.0~${\mu}m$, and which had uncertainties $< 0.1$ in all IRAC-2MASS bands, c.f. Fig.~\ref{fig4}.  
They were identified primarily by their [5.8]-[8.0] color, which is not 
significantly affected by extinction (Flaherty et al. 2006). 
The criteria used were the following:

\begin{eqnarray*}
-0.2 < m_{3.6} - m_{4.5} < 1.0 \mbox{  and  }  -0.2 < m_{5.8} - m_{8.0} < 0.2 
\end{eqnarray*}

\noindent The [3.6]-[4.5] color was also limited to values less than one to
eliminate contamination from protostars which can have [5.8]-[8.0]
close to that of pure photospheres \citep{hart}.

This sample of reddened photospheres was then plotted on the $J-H$
v $H - [4.5]$ and $H-K$ vs. $K-[4.5]$ color-color diagrams to
define where the reddened photospheres fall in the diagrams
(Fig.\ref{fig4}).  Reddening vectors derived from  Flaherty
et al. (2006) were placed on the diagrams so that all of the pure
photospheres were blueward (to the left) of the vectors.  Sources which
were more than $1 \sigma$ redward of the reddening vectors were
selected as having an excess.

Using this analysis on the $J-H$ vs. $H-[4.5]$ diagram, the following
criteria for infrared excess stars was established:
\begin{eqnarray*}
0.912 \times (m_{J} - m_{H} - .6 + {\sigma}_{JH}) + 0.8 + {\sigma}_{H2} < m_{H}-m_{4.5} \\
\end{eqnarray*}
\noindent where 0.912 is the slope of the reddening curve.  There were
72 sources detected using these criteria, 27 of which are not found with the
IRAC color-color diagram. Of the 27, 12 were found to be contaminants.

In the $H-K$ vs. $K-[4.5]$ diagram, the infrared excess stars were identified
using the following criteria:
\begin{eqnarray*}
0.873 \times (m_{H} - m_{K} + {\sigma}_{HK}) + 0.4 + {\sigma}_{K2} < (m_{K}-m_{4.5}) \\
\end{eqnarray*}
\noindent where 0.873 is the slope of the reddening curve.  There were
50 sources detected via this plot, 8 not detected on the IRAC or $J-H$ vs. $H-[4.5]$
color-color diagrams. Three of the eight were found to be contaminants.

There were 24 sources identified as having an excess on the IRAC-2MASS color-color
diagrams that  did not have photometry in all  the IRAC bands and did not show a
[3.6]-[4.5] color excess.  In these cases, the observed infrared excess
resulted from a discontinuity between the 2MASS and IRAC photometry; such
a discontinuity may be the result of time variability or contaminated  
photometry.  These sources were not selected as infrared-excess sources.

\begin{figure}
\plotone{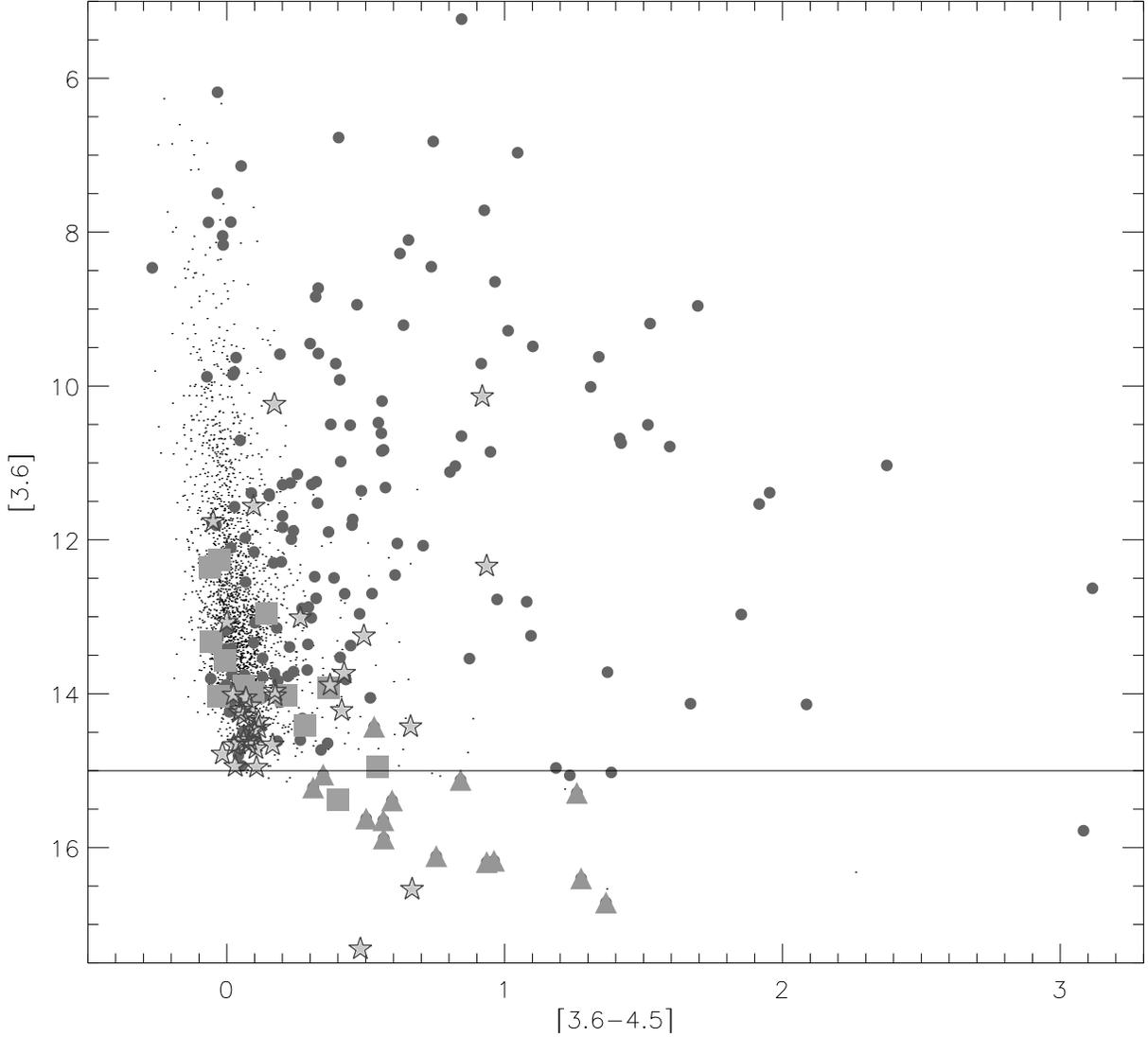}
\caption{ The distribution of YSOs and extragalactic contamination in
the [3.6] vs. [3.6]-[4.5] color-magnitude diagram.  The detections with
photometric uncertainties $< 0.2$ mag in the overlap field are plotted
as dots, with the 229 candidate YSOs overplotted in dark
grey circles.  A majority of the identified YSO sources have
uncertainties $< 0.2$ mag in these two bands. While the star-forming
galaxy contamination (light grey squares) have magnitudes similar to
YSOs, they can be distinguished by their colors.  In contrast, the AGN
(light grey triangles) have colors similar to YSOs, but all lie in the
region below a magnitude of 15 at 3.6~${\mu}m$.  The stars represent
those YSOs that do not possess full four band IRAC
detections; we consider all of these sources with $m_{3.6} < 15$ to be contamination.}
\label{fig6}
\end{figure}

To minimize contamination from AGN, sources which lacked 5.8 and
8.0~${\mu}m$ detections and were identified solely in the IRAC-2MASS
color-color diagrams were rejected if their 3.6~${\mu}m$ magnitude was
fainter than 15~mag. At the distance of Serpens, this magnitude is
well below the cutoff of the Hydrogen-burning limit for 1-3~Myr stars
(12-13~mag; \citet{bar}).  From our analysis of the IRAC color-color
diagram (sec~5.1.1), sources identified as AGN dominate this region
(Fig.~\ref{fig6}).

\subsubsection{IRAC-MIPS Color-Color Diagram}

For young stars with circumstellar disks, the [3.6] - [4.5] vs. [8.0] -
[24.] color-color diagram is very sensitive to the IR excess emission
from dust at ${\sim}$3-5 AU \citep{muz}. It is particularly effective
at identifying stars with little to no mid-IR excess shortward of
8.0~${\mu}m$ due to inner holes in their disks. These transition disks
appear to be in a state of evolution where the inner disk has been
cleared of small dust grains by planets or grain growth \citep{lin}.  We
used the following selection criteria:

\begin{eqnarray*}
0.4 \le m_{8.0} - m_{24} \le 7.0 
\end{eqnarray*}

\noindent A lower limit of 0.4~mag was used to conservatively
eliminate pure photospheres even in the presence of photometric scatter
due to nebulosity and source confusion.  This criteria also excludes
likely extragalactic sources which have colors [8.0] - [24.] $> 7.0$
\citep{muz}.  A total of 75 young stellar objects were identified on
these diagrams, 15 of which were not previously identified in the IRAC
and IRAC-2MASS color-color diagrams. Three of the fifteen were identified 
as extra-galactic contaminants.

\subsubsection{Completeness}

Estimates for the 90\% completeness limit of our {\it Spitzer} photometry were calculated 
via the method of inserting artifical stars into the mosaics and then employing 
our detection algorithms to identify them. The 90\% completeness estimates were 
15.0, 15.0, 14.5, 12.5, 7.5  at 3.6, 4.5, 5.8, 8.0, and 24~${\mu}m$, respectively, 
for the IRAC and MIPS photometry. 

\begin{figure}
\plotone{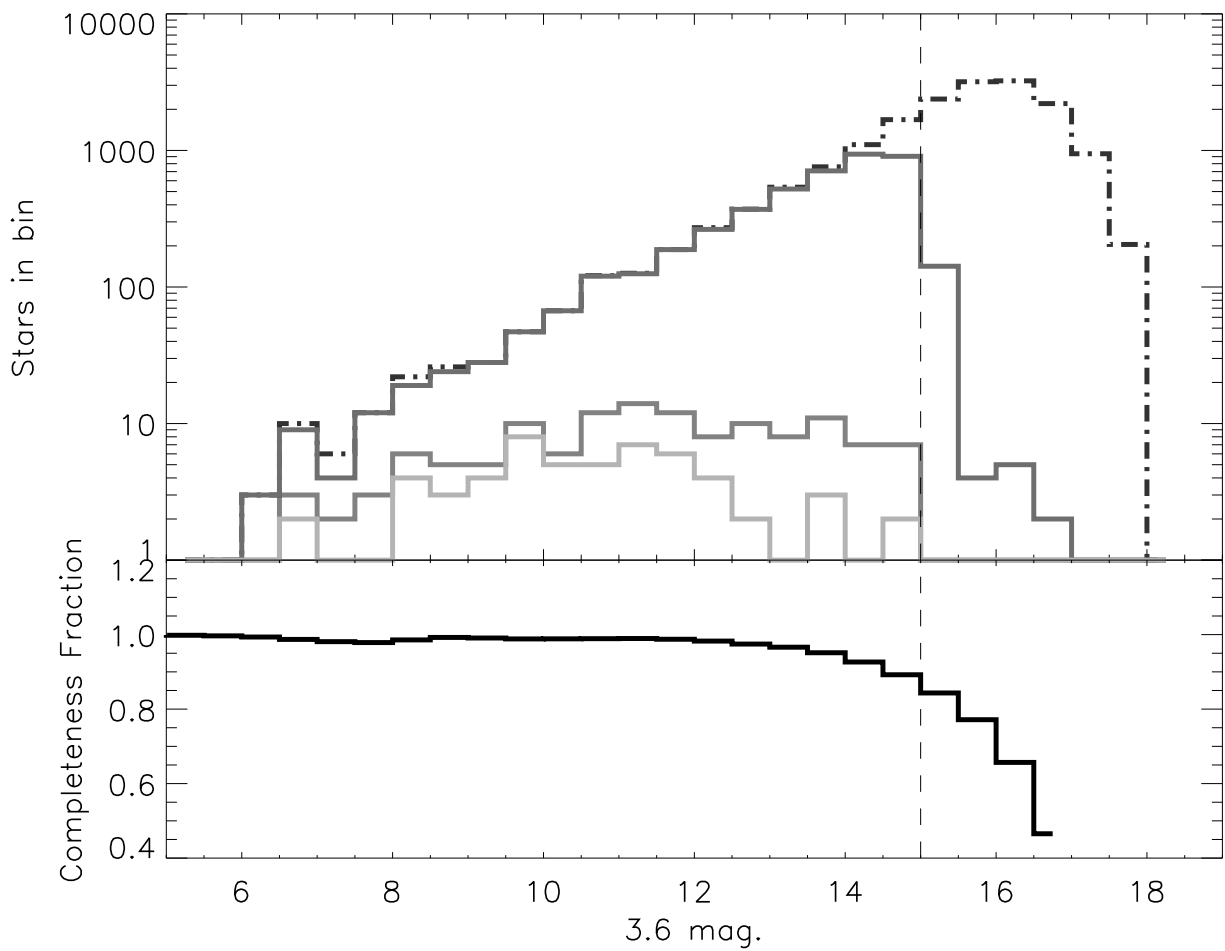}
\caption{ The top panel shows the histogram of all the photometry with a 
detection at 3.6~${\mu}m$, by magnitude, as a black dashed line. Overplotted 
are those sources with detections in the multiple bands needed to place them 
on the color-color diagrams used to select IR-excess sources, as shown by the 
upper gray line.  The lower gray line plots the detections that were selected 
as YSOs in the final catalogue. The light gray line shows the subset of the 
identified YSOs that were detected in X-rays, note that the X-ray observations 
covered a smaller field than the IR observations.
The bottom panel plots the fraction of artifical stars recovered as a function 
of their magnitude for the 3.6~${\mu}m$ IRAC data.
}
\label{figcomp}
\end{figure}

Fig.~\ref{figcomp} plots the histograms and completeness of sources at 3.6~${\mu}m$.
The dashed black line plots the  number of 3.6~${\mu}m$ detections 
by magnitude. 
We have also plotted the number of sources which have sufficient signal to noise 
in the relevant bands that they can be placed on the IRAC, IRAC-MIPS and/or IRAC-2MASS 
color-color diagrams (upper grey line). These are the sources which can be searched for 
IR-excesses.  The final YSO catalogue sources are also plotted (lower gray line), as are 
the YSOs detected in X-rays (lower light gray line).
The ratio of the number of sources with $m_{3.6} \le 15$~mag. that can be placed on the 
color-color diagrams to the total number of sources with $m_{3.6} \le 15$~mag. (and 
correcting this number for completeness by dividing by the fraction of artifical stars 
recovered in each magnitude bin) is $4532/5629$ or 81\%.

The same ratio for detections above 14.5 and 14 magnitudes, the 
percentages are 93\% and 97\% respectively.
These fractions represent lower limits on the completeness of the YSOs, 
as the number of contaminating field stars is rising with increasing magnitude, 
while the number of YSOs does not show an equivalent rise.
Magnitudes of 14.5 and 15 for IRAC 3.6~${\mu}m$, correspond to masses of 
$\sim$0.02~$M_{\sun}$ and $\sim$0.03~$M_{\sun}$ at 1~Myrs, and 
$\sim$0.03~$M_{\sun}$ and $\sim$0.04~$M_{\sun}$ at 3~Myrs \citep{bar}.
This completeness limit is for pre-main sequence stars with IR-excess emission from 
circumstellar disks. Since the fraction of sources with IR-excesses varies with mass and 
age \citep{lad,her}, the completeness with respect to all pre-main sequence cannot be 
estimated with the available data.

\subsection{II: Extremely Red Mid-IR Sources }

Deeply embedded protostars may only be observable at 8.0 or
24~${\mu}m$, and thus not be detected by the previous methods
\citep{gut1}.  In order that these objects are not overlooked, the SED
of each of the 24~${\mu}m$ detections that was not already selected was
visually examined. Three sources were identified at 24~${\mu}m$ only, and
one source was identified at 8.0 and 24~${\mu}m$.  One of sources
detected at 24~${\mu}m$ lies at the position of a known sub-millimetre 
source, SMM1 \citep{tes1}; it is one of the brightest 24~${\mu}m$ sources 
on our image ($m_{24} = 1.6$) and is also seen at 70~${\mu}m$.  
Another source coincides with SMM8 \citep{tes1}; this is detected in 
both 8.0~${\mu}m$ and 24~${\mu}m$. 
It is a fainter source, 11.5 mag at 8.0~${\mu}m$ and 6.3~mag
at 24~${\mu}m$.  One of th esources is near the (sub)mm source SMM3, and is 
probably associated with that source. However, given the lack of clear detections 
in the IRAC-bands and the lack of a clearly identifiable 70~${\mu}m$ counterpart 
in this crowded region, we consider this 24~${\mu}m$ source a tentative YSO.

The remaining source, detected only at 24~${\mu}m$, was
detected in only one of the two epochs of MIPS observations, and is
probably an asteroid.

\subsection{III: X-ray Luminous Stars}
Young stellar objects may also possess elevated levels of X-ray
emission ($ L_{Xbol} \sim 10^3 \times L_{X\odot}$) which can be used
to distinguish them from field stars \citep{fei2,fei3}.  We utilize this
property to identify YSOs that do not have emission from a dusty disk
(evolutionary class III) and would otherwise be indistinguishable from
field stars.  Protostars (class 0/I) and pre-main sequence stars with
disks (class II) with elevated X-ray emission may also be identified.
The {\it Chandra} image of Serpens covered a 17'$\times$ 17' field of view
centred on the cluster, and X-ray data is not available for all the
sources in our catalog, c.f. Table~\ref{tbl2}.

The {\it Chandra} observations detected 88 sources in the region, a mix of
genuine YSOs and AGN. The coordinates of these sources were matched
against the 19,000 IR sources, with 67 matches, or 76\%. 
The lower limit of the X-ray luminosity detectable in Serpens with Chandra can be estimated 
by comparison with the COUP data, see \citet{fei}, as $log_{10}~L_X \approx 28.1$ $ergs$ $s^{-1}$, 
assuming a distance of 260~pc to Serpens and an exposure time of 88~ks. 
We match the X-ray sources with IR analogues to reduce the contamination from AGN; an 
X-ray source with no IR counterpart is assumed to be AGN contamination.  

Of the 67 matches, 40 were IR excess sources previously identified, 3 sources had 
only been detected in some of the infrared bands and consequently
could not be placed on the color-color diagrams used to identify
infrared excess sources.  The remaining 24 sources did not exhibit an IR-excess.
Of the 67 IR matches, 27 of the X-ray luminous stars are not in the list of infrared 
excess sources. Of these 27 sources, five of the sources with {\it Chandra} and 
{\it Spitzer} detections were 
identified as probable AGN using the criteria in Sec~5.1.1.  An additional
two sources which did not show detections in all of the IRAC bands had 
3.6~${\mu}m$ magnitudes fainter than 15; these  seven sources were 
cosidered to be contaminating AGN. In total, 60 YSOs were detected in X-rays, 
20 of which are new to the YSO catalog.
The completeness of the X-ray data can be assessed from Fig.~\ref{figcomp}; 
it should be noted that only 78\% of the YSOs from class 0/I to transistion disk 
lie in the IRX-field, thus the completeness of the X-ray data is limited 
partially by the smaller field of view. 
For $m_{3.6} < 12$ we detect 57\% of the YSOs in X-rays, this 
percentage drops quickly for fainter magnitudes.  This corresponds to a mass of 
0.18~$M_{\sun}$  for an age $\le 3$~Myr and an $A_V = 5$ \citep{bar}.

\subsection{Summary of Identified Objects}

A total of 229 candidate objects were identified in the overlap IR-field through 
the methods outlined above.
A list of 200 candidate members was compiled from the color-color diagrams, 
66 were detected in X-rays in the IRX-field, and three as extremely red mid-IR objects.
After contaminant removal there remained a total of 138, which we consider to be bona 
fide young stellar objects. 
The membership of Serpens is estimated to be more tahn 97\% complete to 14~mag at 3.6~${\mu}m$, or 
$\sim$0.04~$M_{\sun}$ at 1~Myrs \citep{bar}, for sources with mid-IR excess emission.
As will be shown, $\sim$50\% of objects in each evolutionary class are detected in X-rays, 
thus we are likely missing $\sim$21 class III members in the IRX-field. 
By scaling the \citet{har} results to our 0.2~$deg^2$ field, we estimate that there is 
of order 1 AGB contaminant in our catalogue.  We have removed 15 AGN and 13 galaxies with 
strong PAH-emission features, and estimate that $\sim$1 such contaminating object remains 
in the final source list \citep{gut07}.
We estimate that 1 or less of the class III sources may be a dMe contaminant.
A list of the coordinates 137 of the sources identified as young stellar objects in the 
Serpens Cloud Core is given in Table~\ref{tableids} with associated identifiers.  The 
photometry of the sources is given in Table~\ref{tablephot}. These two tables do not list 
the photometry for two sources detected  only by MIPS, SMM1 and SMM3, whose fluxes are listed 
in Table~ \ref{tablesmm}. Since SMM3 is only clearly detected by Spitzer at 24~${\mu}m$, we 
have not included SMM3 in the 138 sources.

\section{\bf Evolutionary Classification}

The evolutionary state of a young stellar object can be inferred from
the {\it Spitzer} mid-infrared photometry \citep{all2}.  The
classification of the Serpens sources was carried out using four
different diagnostics: mid-IR colors \citep{all,meg}, the slope of the
Spectral Energy Distribution (SED) \citep{gut1}, the dereddened SED slope, and the
shape of the SEDs (as ascertained by visual inspection). Each source
was classified as either class 0/I, flat spectrum, class II, transition
disk, or class III \citep[see following subsections for the definition of
these classes]{lad84}.

\begin{figure}
\plotone{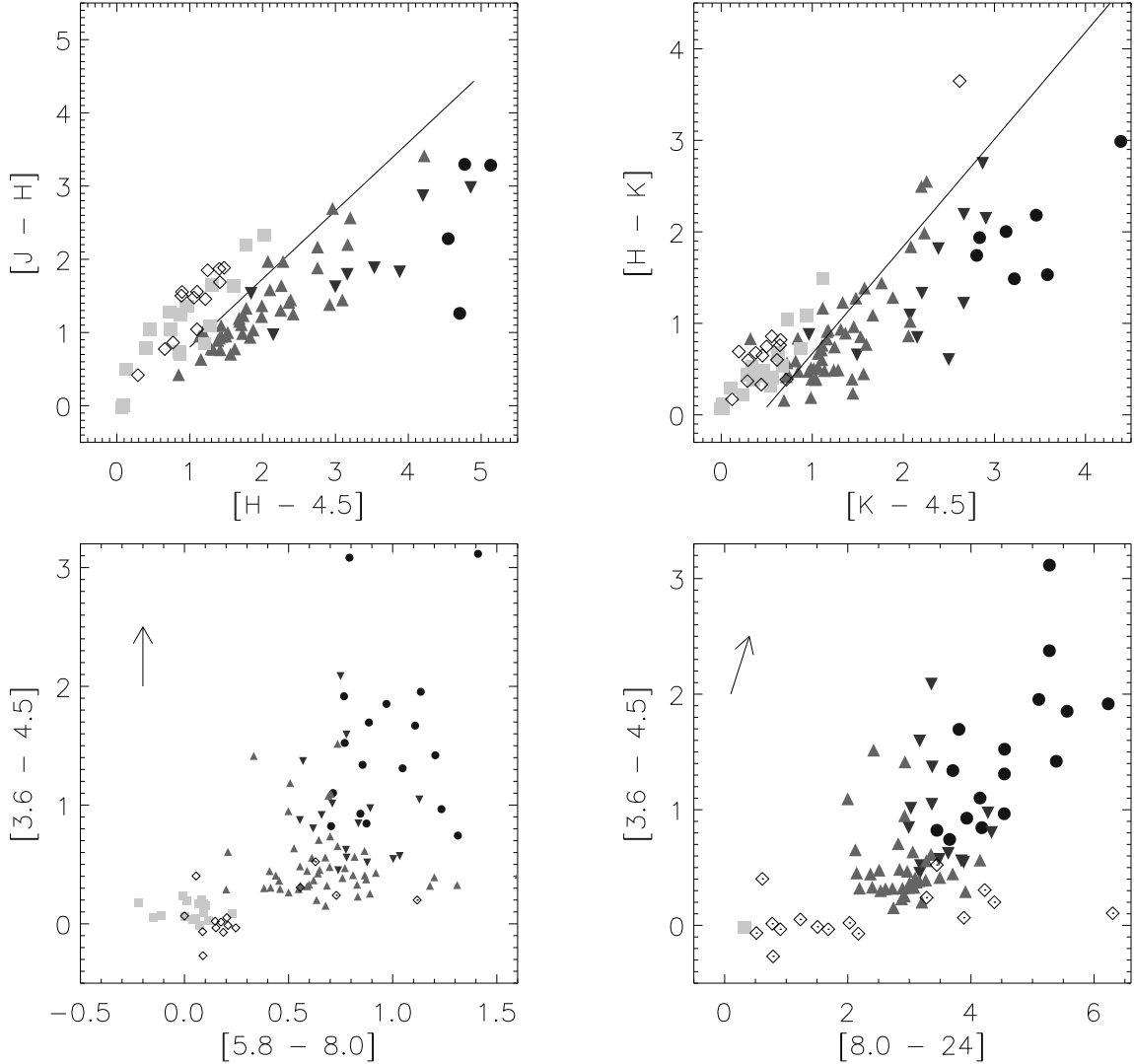}
\caption{The same four diagrams as Fig.~\ref{fig7}, this time
 indicating the positions of the young stellar objects identified in 
 our analysis.  Black circles mark the class 0/I objects, inverted dark 
 grey triangles mark the flat spectrum sources, mid-grey triangles
 indicate the class II objects, while light grey squares show the X-ray
 identified class III sources. The transition disk objects are marked
 with clear diamonds.  
 On the IRAC and IRAC-MIPS color-color diagrams some of the class II 
 and transition disk stars appear to fall in the wrong region of the diagram 
 for their class. 
 An examination of the dereddened SEDs of these sources shows that their 
 positions on the color-color diagrams have been shifted by reddening. 
}
\label{fig8}
\end{figure}

Classification of the sources was carried out by first noting their
locations on the IRAC, IRAC-2MASS and IRAC-MIPS color-color
diagrams and assigning each source a preliminary class (\citet{meg,all,gut,muz}, see Fig.~\ref{fig8}). 
Typically, the IRAC color-color diagram was used
for the initial assignment of the evolutionary class.  An exception
was made in the case of the transition disk objects which do not
always possess an 8.0${\mu}m$ excess and are most clearly identified
by their [8.0] - [24.] color from the IRAC-MIPS diagram.

\begin{figure}
\epsscale{.80}
\plotone{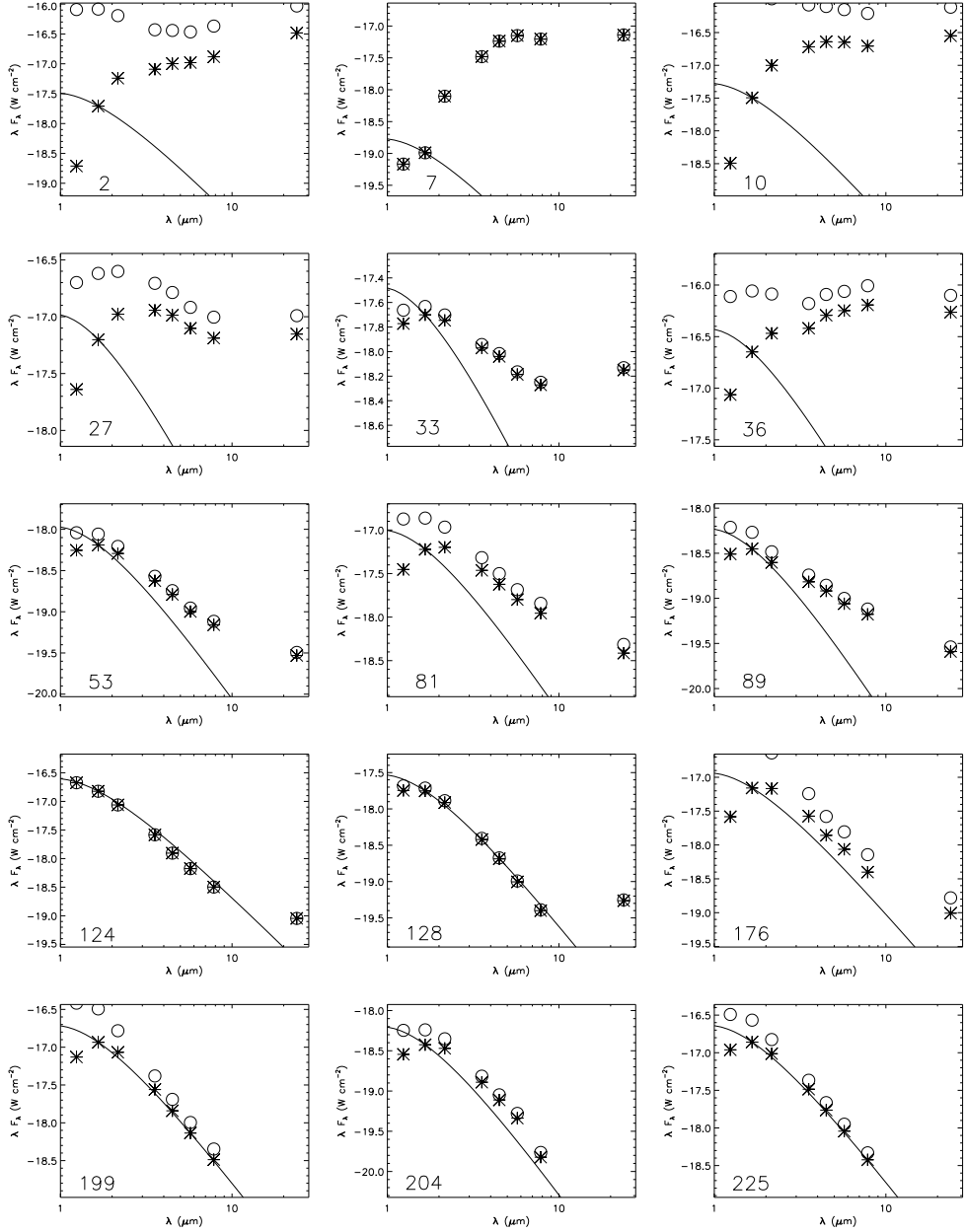}
\caption{SEDs showing three typical examples of each evolutionary class 
in descending order: class 0/I, flat spectrum, class II, transition disks, 
class III. The circles show the measured fluxes, the asterisks the dereddened fluxes. 
The solid line is a 4000~K blackbody scaled to match the $H$-band flux.
}
\label{fig9}
\end{figure}

This was followed by the construction of SEDs for all sources.
Examples of the SEDs for the different evolutionary classes are given
in Fig.\ref{fig9}.  For each SED, a slope $\alpha = dlog(\lambda
F_{\lambda})/dlog(\lambda)$ was calculated.  
The conversion from magnitudes to fluxes
in W~cm$^{-2}$~s$^{-1}$ used the following zero fluxes for
the $J$, $H$, $Ks$, [3.6], [4.5], [5.8], [8] and [24] bands
respectively: $3.13 \times 10^{-13}, 1.13 \times 10^{-13}, 4.28
\times 10^{-14}, 6.57 \times 10^{-15}, 2.65 \times 10^{-15}, 1.03
\times 10^{-15}, 3.02 \times 10^{-16}, 3.80 \times 10^{-18}$.
The slope was calculated
by a least-squares fit over the available IRAC bands, and where
possible IRAC and MIPS. The near-IR magnitudes were not included as
they are most susceptible to extinction.

The Serpens cluster contains many deeply embedded members with
extinctions reaching $\sim$40 $A_V$. For sources
with detections in at least two of the near-IR bands, we have measured
the extinction using a method developed by Gutermuth (2006).
This method uses the extinction law of Flaherty et al (2006) and the
YSO loci  from Meyer et al.  (1997) and Gutermuth (2006)
to determine the extinction and deredden the sources.  The slope
of the SED of the dereddened source was also measured.  

In the majority of cases the SED slope and color-color diagram methods 
agreed (118, 86\%). Where they did not, further examination was undertaken 
to ascertain where the discrepancy arose: 18 members (14\%) were classed 
differently via the IRAC and the dereddened SED $\alpha$ methods. 
Visual inspection of these objects was carried out to better distinguish their
class. 
In 17/18 cases the class derived from the dereddened data was the final choice.  
The remaining two objects were detected at 8.0 and 24~${\mu}m$, and 24~${\mu}m$ 
only, and are coincident with known (sub)mm protostars.

\subsection{\bf Class 0/I Protostars}

Class 0/I sources are protostellar objects surrounded by in-falling
dusty envelopes.  They are characterized by rising SEDs in the
infrared \citep{lad}.  The standard criteria, that class~0 sources with
$L_{submm}/L_{bol} \ge 5 \times 10^{-3}$ \citep{and1}, cannot
be established from the {\it Spitzer} data.  As some known class~0 
sources have been detected with {\it Spitzer} we refer to all
sources showing rising SEDs as class 0/I objects \citep{hat}.  These sources 
were identified in our data by their rising SEDs (${\alpha} > 0.3$; note we used the 
3.6-24~${\mu}m$ slope when available, these may be different to the 3.6-8.0~${\mu}m$ 
slopes listed in Table~\ref{tablephot}), and
by satisfying one of the two following criteria: \citep{all,meg}:
\begin{eqnarray*}
m_{3.6}-m_{4.5} > 0.7  \mbox{   and   } m_{5.8}-m_{8.0} > 0.2  \\
m_{3.6}-m_{4.5} > 0.5  \mbox{   and   } m_{5.8}-m_{8.0} > 1.1  
\end{eqnarray*} 
There can be some overlap between highly reddened class II sources
(stars with disks) and class 0/I objects \citep{whi2}.  To distinguish between these
sources, we used the dereddened SEDs.  Sources that showed dereddened
SEDs that looked like other class II sources  were
deemed class II. These objects are found in the class 0/I region of
the IRAC color-color diagram in Fig.~8. Although all class 0/I objects
might be thought of as class II like objects reddened by their
protostellar envelopes; the envelope also results in the scattering of
a significant component of light in the near and mid-IR \citep{ken,whi1,dop}. 
In addition, there may be a contribution from thermal emission from the inner envelope.
For this reason, class I objects cannot be simply dereddened, 
and the application of our dereddening algorithm results in SEDs which appear 
distinctly different than the class II objects (see Fig.~9).
In total, 22 class 0/I protostars were identified.

\subsection{\bf Flat Spectrum Objects}

Flat spectrum sources possess a 'flat' SED ($0.3 > {\alpha} > -0.3$) \citep{gre},
they do not exhibit the steeply rising SED of protostars (class 0/I), but the
have too much excess for it to arise simply from a circumstellar disk
(class II). Flat spectrum sources are thought to be an intermediate
phase between the class 0/I and class II phase where the central star
and disk are surrounded by a thin infalling envelope \citep{cal2}.
Initially 26 objects were tentatively classified as flat spectrum
sources, using the undereddened $\alpha$, many because they lacked the
necessary near-IR bands to further constrain their classification. This class 
is particularly sensitive to contamination from AGN, as both have similar colors.
Each flat spectrum object was individually examined, and where
possible its dereddened slope was used, to distinguish between class 0/I
($\alpha_{dered} > 0.3$), class II ($\alpha_{dered} < -0.3$), and flat
spectrum, ($ -0.3 < \alpha_{dered} < 0.3$).  We used the 3.6-24~${\mu}m$  
slope for sources with 24~${\mu}m$  detections; in other cases we used 
the 3.6-8.0~${\mu}m$ slope ($\alpha_{IRAC}$) given in Table~\ref{tablephot}.
Finally, the locations of  all the flat spectrum members were plotted on the 
IRAC color-color diagram; these sources delineated the boundary between the 
class 0/I and  II objects (Fig.~8).  In total, 16 flat spectrum objects were identified.

\subsection{\bf Class II}

Class II sources are identified by having an $-2.0 < {\alpha} < -0.3$ \citep{gre,and2}, and 
the following mid-IR colors \citep{all,meg}:
\begin{eqnarray*}
m_{3.6}-m_{4.5} < 0.7 \mbox{   and   } m_{5.8}-m_{8.0} < 1.1  \\ 
m_{3.6}-m_{4.5} < 0.5 \mbox{   and   } m_{5.8}-m_{8.0} > 1.1  
\end{eqnarray*} 
Note that we have lowered the minimum slope from -1.6 to -2 in order to contain evolved or anaemic disks 
found in other star forming regions \citep{lad06,her}.
The class II members were refined by addition of the reddened class II objects initially 
classed as protostars: in total 8 sources (6\%) were reclassified from class 0/I to
class II. Class II sources from the IRAC-2MASS diagrams were distinguished from protostars 
by their slopes in the available IRAC bands. 
In total, 62 class II stars were identified.

\subsection{\bf Transition Disks}
Seventeen sources were plotted that had an excess at 8.0 and/or
24${\mu}m$ but none at shorter IRAC wavelengths.
\begin{eqnarray*}
-0.3 \ge m_{3.6}-m_{4.5} \le 0.1 \mbox{   and   } 0.5 \le m_{8.0}-m_{24} \le 7.0 
\end{eqnarray*}
These sources are considered to be transition disk objects, stars with
a cleared inner disk but retaining an outer disk starting at a few AU 
\citep{muz,mcc}. It is thought that the inner disk might be cleared due to
formation of planets or the agglomeration of the dust particles into
larger mm-sized grains \citep{cal,lin,muz}.  
Those with weaker $m_{8.0}-m_{24}$ excess ($< 1.5$) may be debris disks \citep{muz}.
Identification of these objects by SED slope is unreliable, due to the jump in 
flux at the longer wavelengths (Fig.~9). Visual examination of the SED must be 
carried out to verify that they show excess only longward of 8.0~${\mu}m$. 
They tend to lie near the lower boundary of the class II region on the IRAC
color-color diagram, showing little or no excess on the [3.6] - [4.5]
color-axis and varying degrees of excess along the [5.8] - [8.0]
color-axis (Fig.~8). They are more reliably identified using the
IRAC-MIPS color-color diagram (Fig.~8).  In total we find 17 transition disk 
candidates.   An additional 3 sources were
initially classed by their mid-IR color excesses as Class II sources,
but were identified as transition disk objects from the lack of excess
shortward of 24~${\mu}m$ in their {\it dereddened} SEDs.  These sources
appear in the Class II region in the IRAC and IRAC-MIPS color-color
diagrams.  (Fig.~8).
There is a possible source of contamination in the transition disk sources: 
AGB stars, which have similar colors \citep{blu}, with  $0.5 < m_{8.0}-m_{24} < 1.5$.
This affects eight of the sources, spectroscopy will be needed to determine 
whether these sources are contamination or members.

\subsection{\bf Class III}
Class III objects are pre-main sequence stars which do not have
circumstellar disks detectable in the mid-IR.  They exhibit colors in the color-color diagrams
consistent with reddened photospheres, an SED which approximately
follow a Rayleigh-Jeans law \citep{lad}, and exhibit no excess in the
mid-IR. In total, 21 of the X-ray identified YSOs were found to be class
III.  We cannot distinguish between class III stars without elevated
X-ray emission and field stars in the line-of-sight.  Considering both
the smaller field of view of the {\it Chandra} observations and the fact
that only a fraction of young stars show elevated X-ray emission
(see the X-ray discussion in Sec~\ref{xrayc}), a substantial number of unidentified 
class III objects may exist in the cluster.  Further spectroscopic information 
on stars in the field will be required to positively identify these Class III
sources.

\section{Discussion}

We have identified 138 young stellar objects: 22 class 0/I objects, 16 flat spectrum 
sources, 62 class II objects, 17 transition disks, and 21 class III objects. 
In the following discussion we use this sample to address four topics. 
First, we compare our results to previous studies of the Serpens embedded cluster.  
Second, we identify {\it Spitzer} analogs to sub-millimeter sources previously found 
in this region. Third, we study the spatial distribution of the sources as a function 
of their evolutionary class. Finally, we discuss the X-ray properties of the YSOs
detected in the {\it Chandra} X-ray observavtions.

\subsection{Previous Infrared and X-ray Studies}

Eiroa \& Casali (1992) observed the Serpens cluster in the J, H, K,
and nbL bands and identified 50 young stellar objects in the region.
We match all 50 sources to detections in our overlap region, although
an offset of up to 9$''$ was needed to identify their counterparts in
our data.  Of the 50 sources, 34 have been identified as YSOs in our analysis. 
The remaining sixteen sources were found to have IRAC colors of reddened
stellar photospheres without infrared excesses or to have too few
bands to classify reliably, and none of these sixteen sources showed
X-ray emission.  These sources cannot be established as YSOs by our
criteria.

A previous study by \citet{kaa} using the ISOCAM instrument
on-board ISO, identified 77 cluster members in a 0.13~$deg^2$ region centred 
on the Serpens core, at wavelengths of 6.7~${\mu}m$ and 14.3~${\mu}m$.  
Of these, 70 matched with a {\it Spitzer} detection in the IR-region, with 
the remaining 7 outside of the overlap region but detected in one or more 
{\it Spitzer} bands. 
A maximum offset of 9$''$ was used to match the sources to our data.
Of the 70 sources, 56 are identified as YSOs in our final catalogue, while 
the remaining 14 were found to be reddened stellar photospheres without 
infrared excesses or X-ray detections. These remaining 14 were not identified 
as YSOs in our analysis. Of the YSOs in common with Kaas, two of our class II 
objects are classified as class I by Kaas and one of our class III objects is  
classified as class II by Kaas.

A previous study of the Serpens Core was undertaken in X-rays by
\citet{pre1} and \citet{pre2} with the XMM-Newton Satellite. 
In one observation of $\sim10$~ks and two observations of $\sim20$~ks 
each over and energy range of 0.15-5~keV, 47 sources were detected; 
4 class I, 2 flat spectrum, and 41 class II young stars.
These observations are not as deep as the 88~ks {\it Chandra} integration,
and do not possess as high an angular resolution. We detect a total of
60 X-ray sources in the {\it Chandra} data, 13 more members than the 
Preibisch study.  Twelve of these newly discovered X-ray sources are 
class 0/I or flat spectrum sources, indicating that the {\it Chandra} 
observations are detecting more deeply embedded X-ray sources.

\citet{eir2} describes VLA 3.5cm observations of the Serpens Core,
identifying 22 sources, thirteen of which matched to infrared
counterparts. Nine of these are class 0/I sources including SMM1 and SMM5, 
one is a flat spectrum, two are class II and one is class III.
Three further sources were matched to counterparts detected in one or two bands only. 
The class III object is detected in the  X-ray, thus all thirteen sources have one 
of our two criteria for membership: an infrared excess and/or an infrared counterpart 
(with or without an excess) to an X-ray detection. The remaining six sources without 
infrared counterparts are  most likely background sources which are too faint and/or 
reddened to be detected in the 2MASS or {\it Spitzer} infrared data.
The fluxes of the VLA sources coincident with the protostars detected in the 
(sub)mm are given in Table~\ref{tablesmm}.
Eight of the 13 VLA sources with infrared counterparts have {\it Chandra} X-ray 
counterparts as well; hence, 60\% of the VLA sources are detected by Chandra.
In comparison, in the $\rho$ Ophiuchus core (165~pc), 8/28 or 36\% of the VLA 
sources had both X-ray and infrared counterparts \citep{gag}. 
In the Coronet cluster (140~pc),  9/15 or 60\% of the Radio sources have 
been detected in multi-epoch analysis of X-ray data \citep{for}. 
The measurements are consistent within the statistical uncertanties of the 
ratios, even though the sensitivities vary substantially; 
the X-ray detection rates in the Coronet cluster were found in 20 ks observations, while the 
$\rho$ Oph core observation was nearly 5 times longer.

\citet{gio} have recently submitted the results of the {\it Chandra}
observations of the Serpens cluster.  This work was done in
collaboration with our team, and they have reported the evolutionary
classes listed in this paper.  Table \ref{tableids}, listing the
coordinates and identifiers of the YSOs, provides the cross reference
for the \citet{gio} source numbers in the eighth column.

\citet{har} presents IRAC observations of the Serpens cloud as part of the {\it Spitzer} Legacy project 
"From Molecular Cores to Planet-forming Disks" (c2d; \citet{eva}). 
The surveyed area covers a much larger field of $0.89$~deg$^{2}$ than covered in this paper.
They identified 257 young stellar objects throughout the field, in two main groupings.
One, labeled 'A' in \citet{har}, is the Serpens main core cluster studied in this paper, 
the second region, cluster 'B', being about $1^{o}$ to the south of the main core.
An important difference between this work and that of \citet{har} is the selection of class III objects.  
The \citet{har} criterion for  class III sources required an $\alpha < -1.6$; their class III 
objects are sources with weak infrared emission from a disk that we would classify as 
a weak class II or transition disk object. In contrast, our class III sources are defined as 
lacking in infrared emission yet exhibit detectable X-ray emission.
Bearing this in mind, the numbers of class I, flat spectrum, class II, and class III 
members identified in their larger field were 30, 33, 163, and 31, respectively.

\subsection{Protostars in the Serpens Clusters}

The number of young stellar objects identified is 138.  There are 38
protostars (22 class 0/I, 16 Flat Spectrum) and 100 pre-main sequence
stars (62 class II, 17 transition disks, and 21 class III).
Protostellar class 0/I and flat spectrum sources account for 28\% of the
YSOs detected.  The ratio of protostars to pre-main sequence stars with
disks is 48\%; this high number indicates that Serpens is unusually
rich in protostars, and is in agreement with the 56\% fraction found in 
\citet{kaa}.  Over the entire Serpens cloud, the ratio was found to 
be 38\% \citep{har}.   In comparison, the fraction of protostars to
stars with disks detected with the {\it Spitzer} Cores to Disks (c2d)
survey  was 14\% in the IC 348 cluster and 36\% in the NGC
1333 cluster \citep{jor}, and $\sim$50\% in Chameleon II \citep{por}.

\begin{figure}
\epsscale{1.0}
\plotone{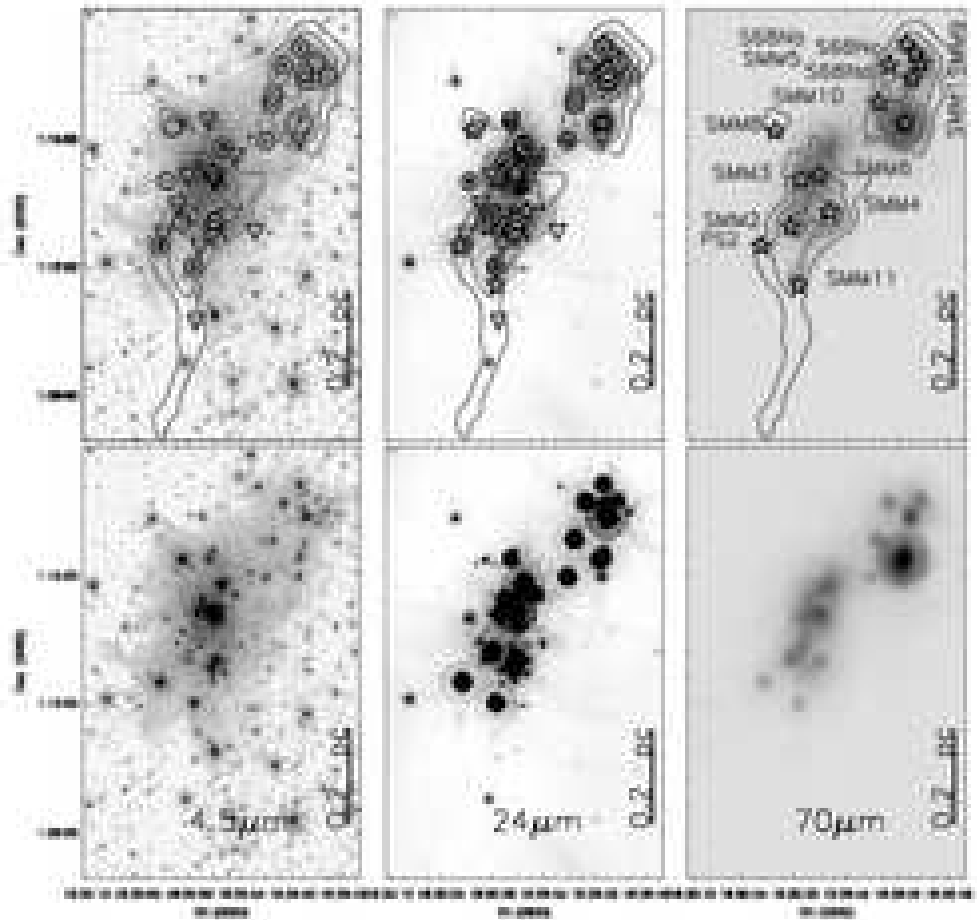}
\caption{ {\it Above}: Three grey-scale images of the Serpens Core, right to left: 
  IRAC Ch.2, 4.5~${\mu}m$, MIPS 24~${\mu}m$, and MIPS 70~${\mu}m$. These are overlaid with 
  contours from the 850~${\mu}m$ SCUBA data \citep{dav1}, at levels of [3\%, 8\%, 20\%, 30\%, 35\%] 
  of the peak flux. The asterisks mark the locations of the identified (sub)mm sources in 
  the region, c.f. Table~\ref{tablesmm}. Class I (circle) and Flat Spectrum (inverted triangle) 
  YSOs are also shown. {\it Below}: The same three greyscale images shown without overlays.}
\label{fig12}
\end{figure}

Previous observations at sub-millimeter and millimeter wavelengths
have identified a number of presumably protostellar sources in the
Serpens cluster.  Studies by \citet{tes2}, \citet{dav1}, and
\citet{wil1} identify fourteen objects detected at wavelengths of 3~mm
or 850~$\mu$m  with flux densities greater than 3 mJy beam$^{-1}$, and 
significance above 5~$\sigma$.   Table~\ref{tablesmm} lists these
objects with their identifiers, coordinates, and fluxes from the above
mentioned papers combined with our {\it Spitzer} photometry.  These
are also shown in Fig.~\ref{fig12}.   We now discuss these sources:

{\bf SMM1:} At the location of SMM1 we find a bright 24 and 70~${\mu}m$
source. Outflow knots associated with SMM1 are visible in our
4.5~${\mu}m$ IRAC mosaic, and diffuse emission is visible in the 8.0~${\mu}m$; 
however, we do not identify a compact source at this location. 
This appears to be a umambiguous example of a Class 0
object too deeply embedded to be detected shortward of 24~${\mu}m$.

{\bf SMM2:} Two {\it Spitzer} identified class 0/I sources are found within $13''$  of the
SMM2 source; these sources appear to be beyond the positional
uncertainties in the 3~mm data ($\sim6''$) and are not coincident with SMM2.  
There is also 70~$\mu$m emission in the region around SMM2, it is not 
clear which of the three sources (SMM2 and the two class 0/I) contribute to 
70~$\mu$m emission.  Interestingly, there is a faint slightly extended 4.5~$\mu$m 
source toward the position of SMM2.  This suggests that the SMM2 source 
may be a protostellar source with associated infrared emission, 
although the lack of  24~$\mu$m emission is puzzling.

{\bf SMM3:} There is a 24~$\mu$m source within $2''$ of the location of SMM3, 
which we tentatively assign to it.  While there is extended 70~$\mu$m 
emission in the region of SMM3, no point source could be photometered due to flux 
contamination from nearby bright sources.  While there is faint emission 
towards this source in the IRAC bands, it is confused with image artifacts from 
neighboring bright sources.

{\bf SMM4:} There are no sources identified in the IR coincident with  
SMM4, but two small areas of nebulosity can be seen in the 4.5${\mu}m$  
image. Again,  this suggests that a protostellar source may reside 
in SMM4, but if so, there is a lack of bright 24~$\mu$m and 70~$\mu$m 
emission from the source.

{\bf SMM5, SMM9, S68Nb, S68Nc and S68Nd:} These five sources are
associated with the group of four class 0/I objects on the
northwestern side of the cluster.  In addition, the outflow jet associated with
object S68Nc is visible in the IRAC bands.  This region has previously
been considered devoid of more evolved pre-main sequence stars, however
we have identified two class II members in the vicinity.

{\bf SMM6, SMM8, SMM10, PS2:} There is a flat spectrum source at the
position of SMM6 and class 0/I objects at the positions of
SMM8, SMM10 and PS2. Recent work by \citet{hai2} identified SMM6 
as the primary  component of a double system. We do not see the 
companion in our data, as it is contaminanted by the flux from SMM6 itself.

{\bf SMM11:} While there is no IR counterpart at the position of SMM11 
from \citet{tes1},  a class 0/I object $\sim20''$ north of its 
position is detected out to 70~$\mu$m.  We also note that the peak of SMM11 in 
the SCUBA 850~${\mu}m$ data is $\sim10''$ southeast of our class 0/I 
source and $\sim8''$ northwest of the Testi source. While it is possible 
that these are three separate objects, it seems unlikely; the position of this 
source needs to be re-examined in subsequent submillimeter observations.

\begin{figure}
\plotone{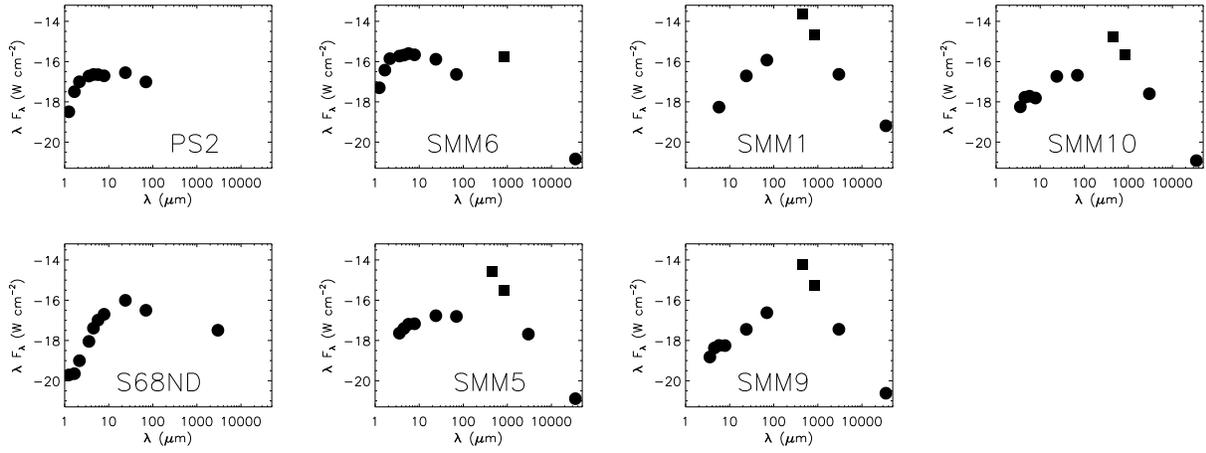}
\caption{ Spectral Energy Distribution of seven cluster members detected at 70${\mu}m$, 
 with 3~mm and 3.5~cm data points plotted where available \citep{eir2,wil2}. The black 
 squares indicate the peak flux per beam at 450 and 850~${\mu}m$ from \citet{dav1}.
}
\label{figsmmseds}
\end{figure}

Of the fourteen (sub)mm objects, seven correspond directly to sources 
detected in the IRAC bands, three are detected only with MIPS, and four are not 
detected at $\lambda \le 25~\mu$m. 
With the exception of SMM11, all of the (sub)millimeter sources are coincident with
70~$\mu$m emission; however, the low angular resolution of the
70~$\mu$m map ($16''$) makes it difficult to extract the
70~$\mu$m photometry for an individual source.  
In seven of these sources, the 70~$\mu$m
flux can be extracted; these fluxes are tabulated in Table~\ref{tablesmm} and the 
resulting SEDs are shown in Figure~\ref{figsmmseds}.
For the remaining sources, the 70~$\mu$m emission
is too confused to photometer accurately.  
Three of the fourteen sources have X-ray detections in our data: SMM5, SMM6,
and S68Nb.  These appear to be Class I sources in that they show
bright emission shortward of 24~$\mu$m; there is no detectable X-ray
emission toward the sources with weak or no detected
emission shortward of 24~$\mu$m, i.e. the probable Class 0 sources.

\subsection{ Spatial Distribution}

The spatial distribution of young stellar objects in a cluster gives
insight into the fragmentation processes leading to the formation of
protostellar cores and the subsequent dynamical evolution of the stars
as they evolve from the protostellar to the pre-main sequence stage
\citep{all2}.  Recent work has shown that in many clusters the
sources trace the underlying molecular gas distribution \citep{gut}. 
\citet{kaa} made the first maps of the distribution of class II and 
class I objects in Serpens, demonstrating that the class I sources 
were significantly more clustered than the class II sources.

\begin{figure}
\plotone{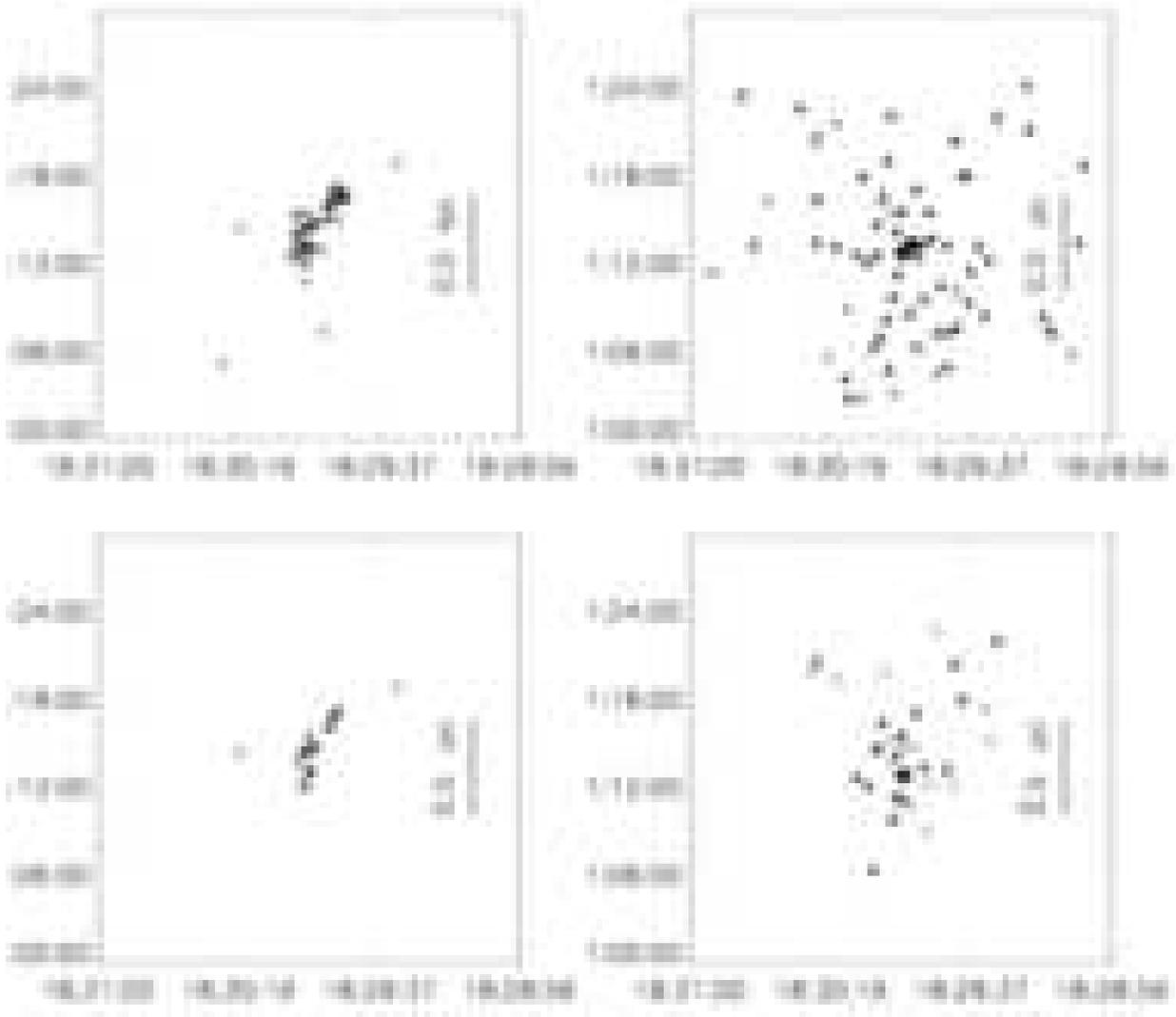}
\caption{ {\it Above}: The spatial distribution of the various classes of YSOs
  in the Serpens cloud core. {\it Upper Left}: Class I and flat spectrum sources, in 
  circles and inverted triangles respectively. {\it Upper Right}: Class II (triangles) and transition 
  disk members (diamonds). {\it Lower Left}: X-ray selected sample of class I and flat spectrum 
  objects. {\it Lower Right}: X-ray selected sample of the class II, transition disk, and class III 
  members (shown by squares). The dense clustering of the class I and flat 
  spectrum sources can be seen, over the more widely dispersed class II and later members. 
  Note: the extent of the IRX-field is smaller than the IR-field, as shown in Fig.~\ref{fig2}.}
\label{fig13}
\end{figure}

Fig.~\ref{fig13} shows the distribution of young stars as a function of their
evolutionary class.  This is shown for the sample of IR-excess
stars and the sample of X-ray luminous stars.  The most significant
difference is between the class 0/I and flat spectrum sources, which
are concentrated in a narrow filament in the center of the cluster,
and  the class II sources.  Although there
is a peak in the class II density in the center of the cluster, the
majority of the class II sources, as well as the transition disk and
class III sources, are found in an extended halo surrounding the
protostars.  

The protostars are coincident with the dense molecular ridge 
mapped by the 850~${\mu}m$ SCUBA map of \citet{dav1,dav2}.
The class 0/I and flat spectrum sources concentrated in  two groups coincident  
with the two dominant molecular gas clumps (Fig.~\ref{fig12}), confirming the previous 
work of \citet{tes2} and \citet{kaa}. These are coincident with the column density peaks 
of the 850~$\mu$m map and have size 0.2~pc in diameter.  The northern group contains 
9 Class 0/I and 1 flat spectrum source, seven of which are found in a region only 0.1~pc 
in diameter, and 4 class II sources.  
The southern group contains a mixture of 12 class 0/I, 10 flat spectrum, 
16 class II, 1 transition disk, and 5 class III sources in a $\sim0.2$ pc diameter region.
In the southwestern quadrant of this grouping is a small wishbone shaped
sub-grouping. This sub-grouping of stars contains eight stars in an area
0.1~pc in diameter; including three class 0/I objects, one flat
spectrum source and four class II sources.  The higher fraction of class II sources
 in the southern group suggests that it is more evolved than the northern group.  
Finally, the X-ray data shows a grouping of four class III objects in the
southwestern quadrant of the cluster within a  0.1~pc diameter region.  This
grouping is unusual in that it is the only apparent group of class III
objects; all of the remaining class III objects are scattered around
the cluster.  If this is a bona fide group and not a chance alignment, 
it is of great interest why these objects have dispersed their circumstellar material.

\begin{figure}
\plottwo{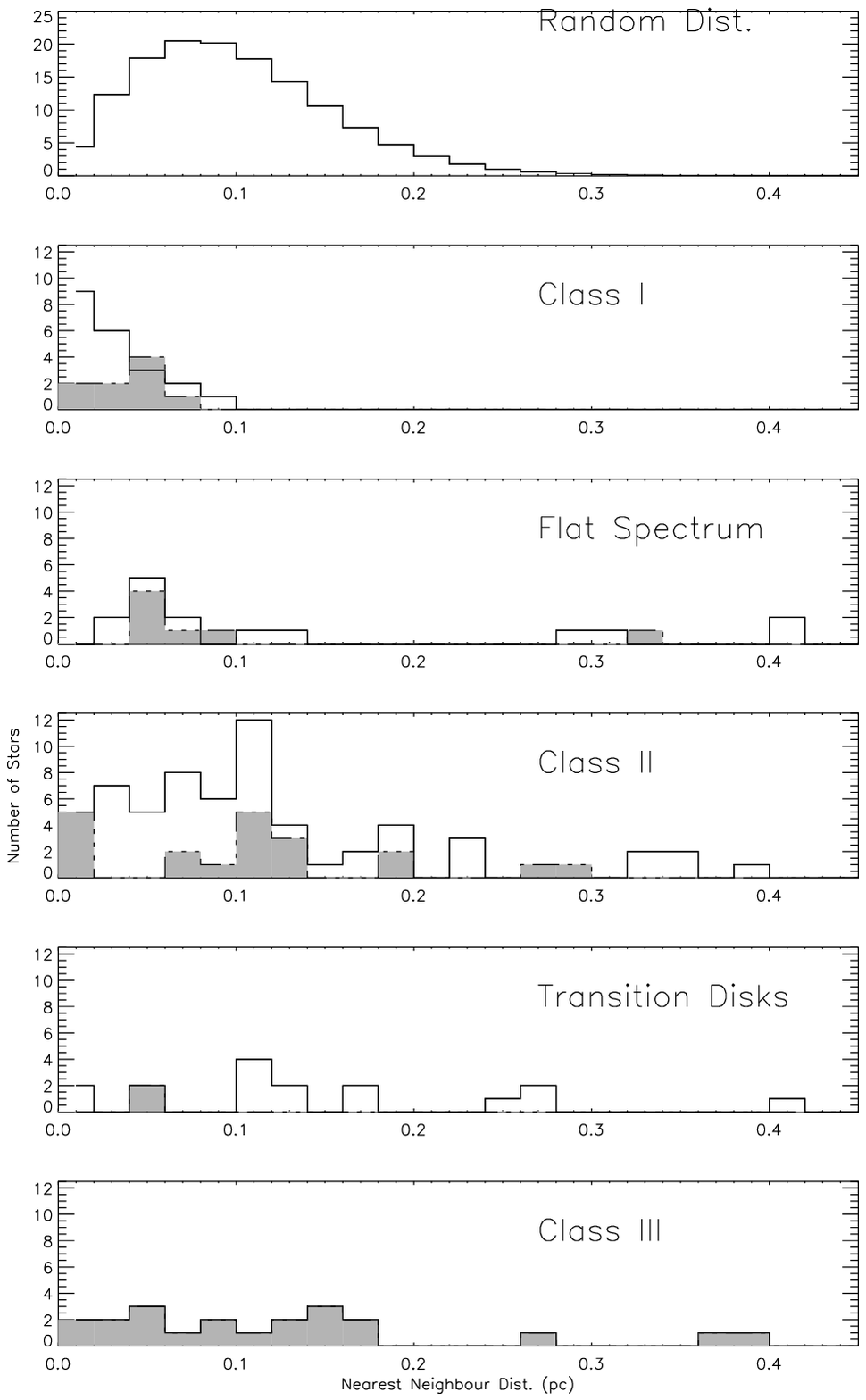}{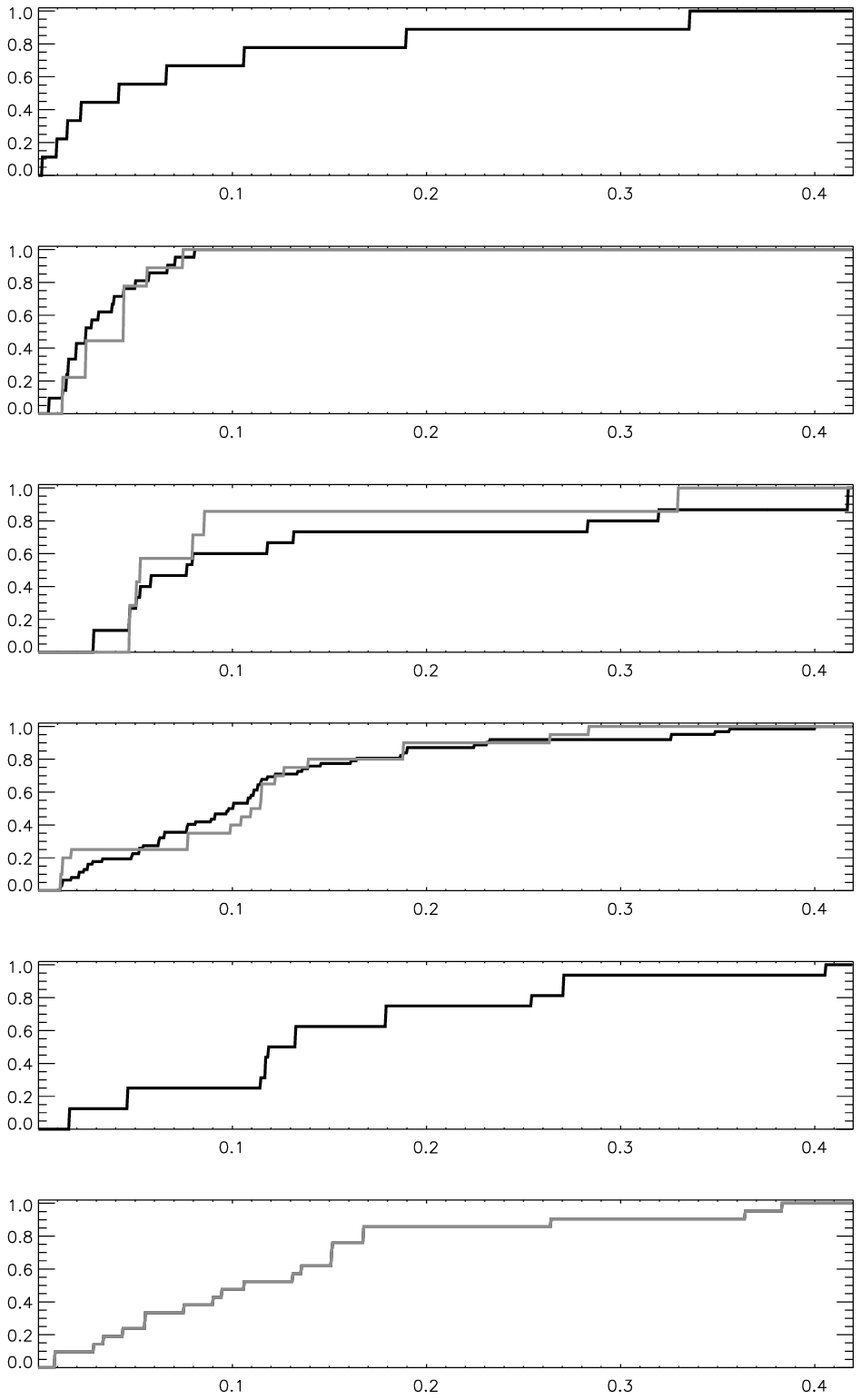}
\caption{ {\it Left:} Differential distribution of nearest neighbour distances by class for the
 identified YSOs (solid line), the shaded region gives the differential distribution of 
 nearest neighbour distances by class for the X-ray selected sample of YSOs. The protostellar
 sources are densely clustered, with a wider distribution for the stellar sources. 
 The top plot is the nearest neighbour distances for a random distribution.
 {\it Right:} Cumulative distribution of nearest neighbour distances by class for the identified YSOs, 
 the grey line gives the X-ray sources. The top plot is the cumulative plot of a random distribution. 
 A Kolmogorov-Smirnov probability shows the class I distribution to be different from the class II or 
 class III distributions, and that the class II and class III distributions are statistically similar to 
 the random distribution.}
\label{fignn}
\end{figure}

The spatial distribution of the cluster sources was examined for each
of the five evolutionary classes using a nearest neighbour technique.
The nearest neighbour distance is the projected distance to the
nearest YSO of the same evolutionary class, using the adopted distance
of 260 pc.  Figure \ref{fignn} ({\it left}) shows the distribution of
nearest neighbour distance between the members of each class. 
The class 0/I sources are by far the most densely clustered, with a
median separation of 0.024~pc. In comparison, the median nearest neighbor
distances of the remaining evolutionary classes are significantly
greater: flat spectrum - 0.079~pc, class II - 0.097~pc, transition disks -
0.132~pc and class III - 0.131~pc.  The median distances for the flat
spectrum and transition disk are biased to higher values by the small
numbers (and hence low densities) of these sources.  The class III
sources mean distance may be biased to a lower values by the limited
field of view of the {\it Chandra} data.  
These scales are smaller than the 0.12 and 0.25~pc separations for class I 
and II previously derived from ISOCAM observations \citep{kaa}, which were 
limited by the lower angular resolution of ISO.
Also plotted in Figure~\ref{fignn}
are the cumulative distribution curves for the YSOs by class, showing 
how densely the sources are distributed over the field.  A
Kolmogorov-Smirnov (K-S) test was performed to ascertain the probability 
that they are derived from the same parent distribution.  The Class I sources are
dissimilar to each of the other classes, reflecting their more highly
clustered nature (Table~\ref{tprob}).  The distributions of the four remaining classes are
statistically indistinguishable (Table~\ref{tprob}).  

For each evolutionary class, 10K random distributions of stars were  
generated, with the number of stars equal to the number of objects 
in the given evolutionary class.  For the class III sources, the
randomly distributed stars were constrained to fall within the IRX-field, 
for all other classes, the random distribution covered the region of the 
IR-field that contained 90\% of the sources. The resulting nearest neighbor 
distributions of the observed YSOs and the random distributions were
compared for each class using the K-S test.  The probabilities that the 
random and observed distributions were drawn from the same parent 
distribution were calculated for all 10K distributions; the mean values 
of the probablilities for each class are listed in Table~\ref{tprob}.

Our results indicate that the median projected spacing of protostars in
Serpens sources is only 5000 AU, and the spacing is as close as 2000 AU in
certain regions such as the wishbone and the northern group.  The 
actual physical spacings are probably larger; if we assume the sources are randomly
distributed relative to each other, the median spatial separation
between sources would be $1.27 \times 5000 = 6350$~AU.  
If we assume that each protostar will form an 0.5~$M_{\sun}$ star, then for dense core densities of 
$2.5 \times 10^4$ and $6 \times 10^5$~cm$^{-3}$ \citep{jij,olm}, such a protostar would have
to accrete from a volume 12,000 and 4000 AU in diameter, respectively.  Thus,
assuming volume densities typical of dense molecular cores, the
accretion volumes would be densely packed if not overlapping.

Is the observed spacing consistant with Jean's fragmentation?  The
Jean's length is given by \citep{jea}: 
$\lambda_J = 0.21~pc~(T/10~K)^{1/2}~(n_{H_2}/10^4 cm^{-3})^{-1/2}$.  
If we set the Jean's length to be $1.27 \times 0.024$~pc,  adopt 
temperatures of 12 and 17~K  \citep{jij},  and solve for the density, 
the resulting densities are $6 \times 10^5$~cm$^{-1}$ and $8 \times 10^5$~cm$^{-1}$.  
The observed gas densities range from 
$2.5 \times 10^4$~cm$^{-3}$ to $6 \times 10^5$~cm$^{-3}$
\citep{jij,olm}.  Thus, the observed median separation is
consistant (i.e. within a factor of two) with Jeans fragmentation
given the highest observed densities in Serpens. 
 At these highest observed densities, the
spacing would also be approximately equal to the radius of a critical Bonner-Ebert sphere
($R = 0.22~pc~(T/10~K)^{1/2}~(n_{H_2}/10^4 cm^{-3})^{-1/2}$) \citep{ebe,bonner}.  
In either case, the spacing of objects at the Jeans spacing and
Bonner-Ebert radius further indicate that the volumes of gas accreting
onto the protostars are tightly packed or overlapping. We note that
unlike the case of \citet{tei}, there is not a well defined
peak in the distribution of nearest neighbor separations; many objects
are found at separations much smaller than the median separation. 
This is in keeping with \citet{you} who resolved one of the 
Teixeira et al. sources into a multiple system.

The close spacing of the protostars raises the possibility
that competitive accretion is taking place.  In competitive accretions
models, several protostars accrete from a common reservoir of gas.
These objects compete for gas from the reservoir; this
process can lead to a distribution of masses similar to the initial
mass function \citep{bate2005}.  In models of competitive accretion, 
protostars accrete mass through a Bondi-Hoyle accretion at a rate of  
\begin{eqnarray*}
\dot{M_{\star}} = 4 \pi \rho \frac{(G M_{\star})^2}{(c^2+v^2)^{3/2}}
\end{eqnarray*}
where $\rho$ is the volume density
of the gas, $M_{\star}$ is the instantaneous mass of the accreting
protostar, $c$ is the sound speed of the gas and $v$ is the velocity
of the star relative to the gas \citep{bon,bonnell2006,shu1992}.
\citet{wil2} measured the velocities of 5 of the protostars
in the northern group by using detections in the $N_2H^+$~($1\rightarrow 0$) 
line conicident with the protostars.  The RMS 1-D velocity
dispersion of these protostars is 0.26~km~s$^{-1}$, implying a 3-D 
velocity dispersion of 0.45~km~s$^{-1}$.  
Assuming a density of $6 \times 10^{5}$~cm$^{-3}$, a velocity of 0.45~km~s$^{-1}$,
 and a kinetic temperature of 
17~K  (implying a sound speed of 0.22 km~s$^{-1}$), 
for a protostar with a mass of 0.1~M$_{\odot}$, the Bondi-Hoyle accretion  
rate is $8 \times 10^{-7}$~M$_{\odot}$  per year, and for a protostar with a mass of  
0.5~M$_{\odot}$, the accretion rate is $2 \times 10^{-5}$~M$_{\odot}$ per year.   
Given these accretion rates, adopting a protostellar
lifetime of 300,000 \citep{hat},  and considering that the  
rate will increase
as the object grows in mass, the young protostars in the northern   
group could  accrete a significant portion of their ultimate mass  
through Bondi-Hoyle accretion. Consequently,  competitive accretion  
remains a viable process for the formation of stars in the Serpens  
cluster.

Finally, the close spacing and velocities of the protostars in the northern 
group suggest
that their envelopes could impinge on one another.
We note that the 3~mm
continuum observations ($5.6''$ beamsize) of \citet{wil2}
did not resolve the the sizes of the protostars in the northern 
group of Serpens.  
This does not rule out our previous estimates that the accretion 
volumes extend 5000 to 10,000 AU;  the BIMA
measurements may only be sensivite to dense inner regions of the
protostars.  However, for the present analysis, we use a size of
1000~AU, assuming that the protostars are spheres with diameters just
below the angular resolution of the BIMA observations.  
We follow the analysis in \citet{gut}, adopting a mass of 0.5~M$_{\odot}$, a velocity of
0.26~km~s$^{-1}$ and a peak density for the northern group of seven protostars in a
volume 0.1~pc in diameter.  The resulting time between collisions for a given 
object is 200,000 years, less than the protostellar lifetime.
Consequently, if the objects in this group remain in bound orbits
within this region for 300,000 years, the typical protostellar lifetime \citep{hat}, 
all of the protostars could experience collisions between their envelopes and the 
envelopes of other protostars in the group.

In summary, Jean's length, Bonner-Ebert sphere and simple volumetric
arguments all suggest that the accreting envelopes around the
protostars in Serpens are often tightly packed, if not overlapping.
We find that for the conditions present in the northern group, the
protostars may accrete a significant amount of their mass through
Bondi-Hoyle accretion from a common envelope, and competitive 
accretion may occur within the group.  
Furthemore, collisions  between the sources are likely.  
In total, our analysis suggests that the interactions between protostars 
can play an important role in the formation of stars within the dense 
groups observed in Serpens.

\subsubsection{The Substellar Candidates and their Spatial Distribution}

Using the models of \citet{bar} we calculate the magnitude of the hydrogen burning limit at 
K-band to be 12.5 mag for an assumed 1~Myr old cluster at a distance of 260~pc to Serpens.
Of the 138 sources in Serpens, 34 have K-band detections fainter than 12.5 magnitudes, 2 of which 
are class 0/I and one flat spectrum. Since we cannot measure the dereddened photospheric 
luminosities of the class 0/I and flat spectrum objects, these sources are discounted, leaving 
31 candidate (23 class II and 8 class III) brown dwarfs members.  \citet{har} has also identified 
numerous substellar candidates in the Serpens 'B' cluster to the south of the core.
We cannot rule out that these sources are older objects or objects that appear underluminous 
for other reasons \citep{pet,sle}.   Searches for brown dwarfs in other regions have shown that 
infrared spectroscopy is essential to confirm that faint objects are protostellar in nature \citep{luh2}.
 Assuming that the 23 class II objects were to be positively identified as substellar, this would 
lead to an upper limit of 23/138 or 17\% of the total YSO population being substellar.
This would be somewhat higher than the substellar fractions found in Chameleon 1 (5\%) and 
IC348 (8\%) \citep{luh1,luh3}, though in agreement with \citet{kaa}, who estimate 20\% of Serpens 
sources are substellar.

\begin{figure}
\plotone{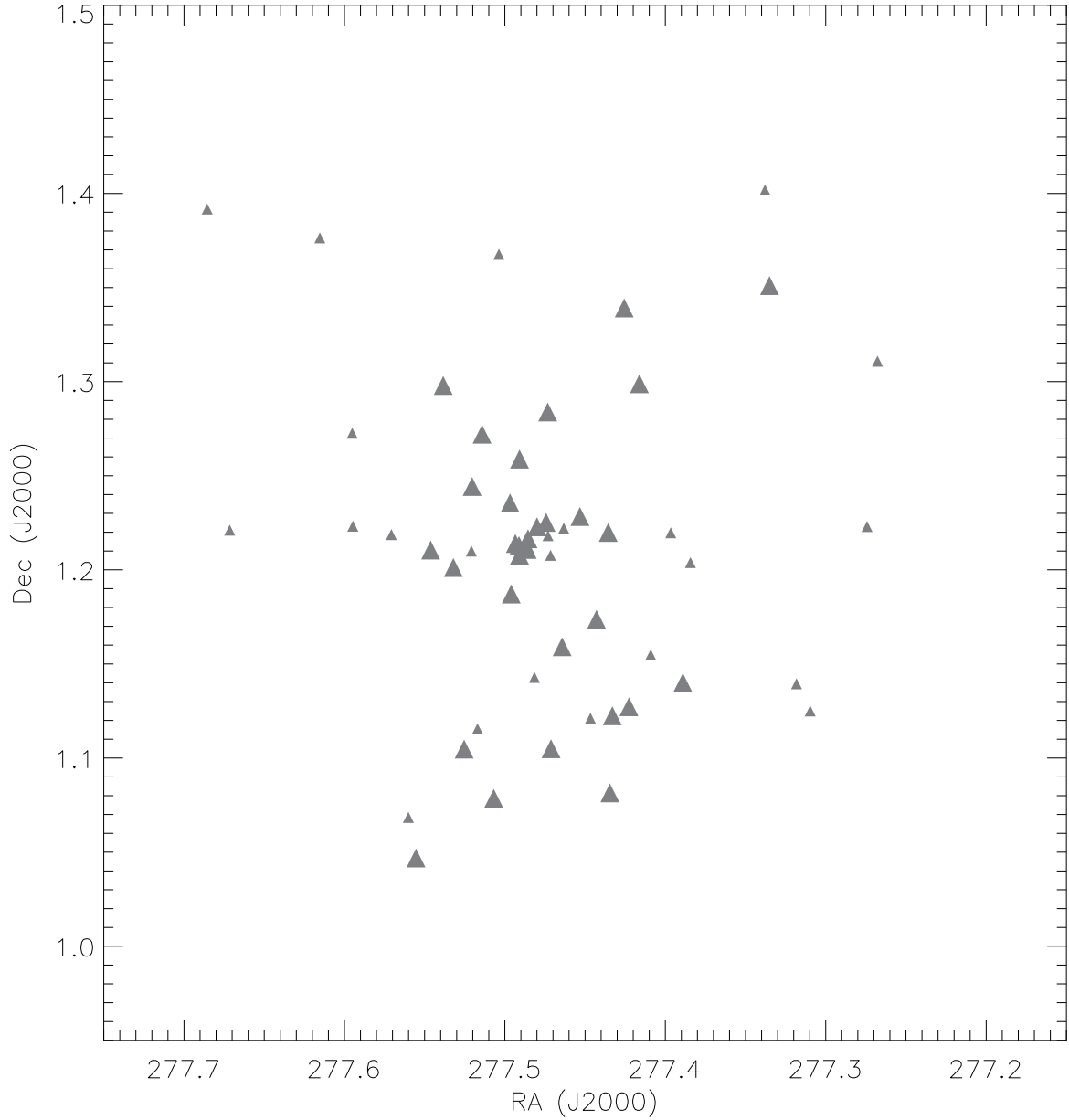}
\caption{Spatial Distribution of the class II sources in the Serpens cluster. The larger
         symbols represent the sources with dereddened K-band magnitude brighter than 12.5, 
         the smaller represent the sources with K-band magnitude fainter than 12.5. The two 
         have similar distributions, indicating a similar formation mechanism for the two groups. }
\label{figbdc}
\end{figure}

Interestingly, the spatial distribution of the $m_K >$ 12.5 class II objects appears 
more extended that the brighter class II (Fig.~\ref{figbdc}).  In particular, although 
smaller in number, the most distant objects north, west, and east of the 
core of the cluster have $m_K >$ 12.5.  This may be biased by the high reddening 
in the centre of the cluster which would preclude detections of faint objects in the 
near-IR. 
Using deeper near-IR photometry of the center of the cluster  \citet{kaa2} also found that 
sources with dereddened $m_K >$ 12.5 were non-centralised.
Spectroscopy of the fainter objects is needed to confirm their substellar nature.

\subsection{X-ray Characteristics \label{xrayc}}

In recent years many studies have been undertaken to investigate the emission properties 
of pre-main sequence stars and protostars in the higher energy X-ray region of the spectrum
\citep{wol,get}.
In developed, hydrogen burning stars, X-ray activity arises from magnetic fields generated 
as a result of shear between the core radiative zone and the outer convective zone. 
The process behind the generation of the highly increased levels of emission in young stellar 
objects remains uncertain; suggested causes being magnetic disk-locking between the star and 
disk \citep{hay,iso,rom}, accretion onto the star \citep{kas,fav1,fav2}, and alternative 
dynamo models for coronal emission  \citep{kuk,giam}. 
In main sequence stars, there is a trend of decreasing X-ray activity with age;  this trend 
has not yet been conclusively shown for pre-main sequence stars.
{\it Chandra} has recently examined the X-ray properties of the Orion Nebula Cluster (ONC), 
in a project known as the {\it Chandra} Orion Ultra-Deep Project (COUP). 
Although the Serpens sample is much smaller than the COUP survey, the analysis of Serpens 
has the advantage of excellent 3.6-24~${\mu}m$ photometry; observavtions longward of 
4.5~${\mu}m$ are difficult in the center of the ONC due to the bright infrared emission from the 
molecular cloud.   This photometry allows us to accurately ascertain the evolutionary class of 
each object, and study the dependence of the X-ray properties with evolutionary state.
Only those detections with X-ray count in excess of 100 were considered when examining 
emission properties to ensure reliable estimates.

Of the 138 identified members, 60 were conclusively shown to exhibit  X-ray emission.
Nine of the X-ray detections are associated with Class 0/I sources (43\% of class 0/I's in IRX-field), 
a further eight with flat spectrum members (56\%). Twenty detections have class II 
counterparts (45\%), while two of the transition disks are detected in X-rays (25\%).
The remaining 21 sources did not exhibit an IR-excess and were identified as a class III YSO 
solely on their X-ray emission; hence, 21/138 or 15\% of the sources were identified solely by Chandra.
There appears to be no evidence that the detection rate depends on evolutionary class;  
within 1~$\sigma$, the detection rates of class 0/I, flat spectrum, and class II sources are identical.  
The detection rate may be largely a function of the sensitivity  of the {\it Chandra} observations; 
the majority of T Tauri stars have $log(L_{x}/L_{bol})$ $> \sim$ $-5.0$ \citep{pre4} while our 
sample is complete to $log(L_{x}/L_{bol})$ $> \sim$ $-4.5$.   Since we cannot identify class III 
sources without X-ray emission, we cannot establish the fraction of class III sources detected in 
X-rays.    
As we will argue subsequently, the X-ray properties of the class II and lII objects appear to be 
indistinguishable.  Assuming that this similarity holds for the fraction of detected sources, then 
the fraction of classs III sources detected is the same as that for the class II objects, and 
we estimate the total number of class III objects to be $\sim46$.  
If, further, we take into account that the IRX-field contains only 71\% of the class II population, 
and we assume that the class III sources show a similar spatial distribution, then the total 
population of class III objects may be as high as 65.

The ratio of the number of X-ray detected pre-main sequence stars with disks (class II and transition disk) 
to the total number of X-ray detected pre-main sequence stars (Class II, Class III, and transition disk) 
is found to be 22/(22+21) = 51\%.
The number counts method of \citet{gut} was not applicable here due to the high density of background stars.  
These values are comparable to the disk fraction of a $\sim$2~Myr old  cluster \citep{her}; 
consistent with the suggestion of \citet{kaa}  that the class II population of Serpens is 2~Myr old.  
However, this older age seems odd considering the large number of protostars in this region.
In a future paper, we will examine the positions of these objects on the H-R diagram to better 
determine the age of the Serpens cluster.

\begin{figure}
\plotone{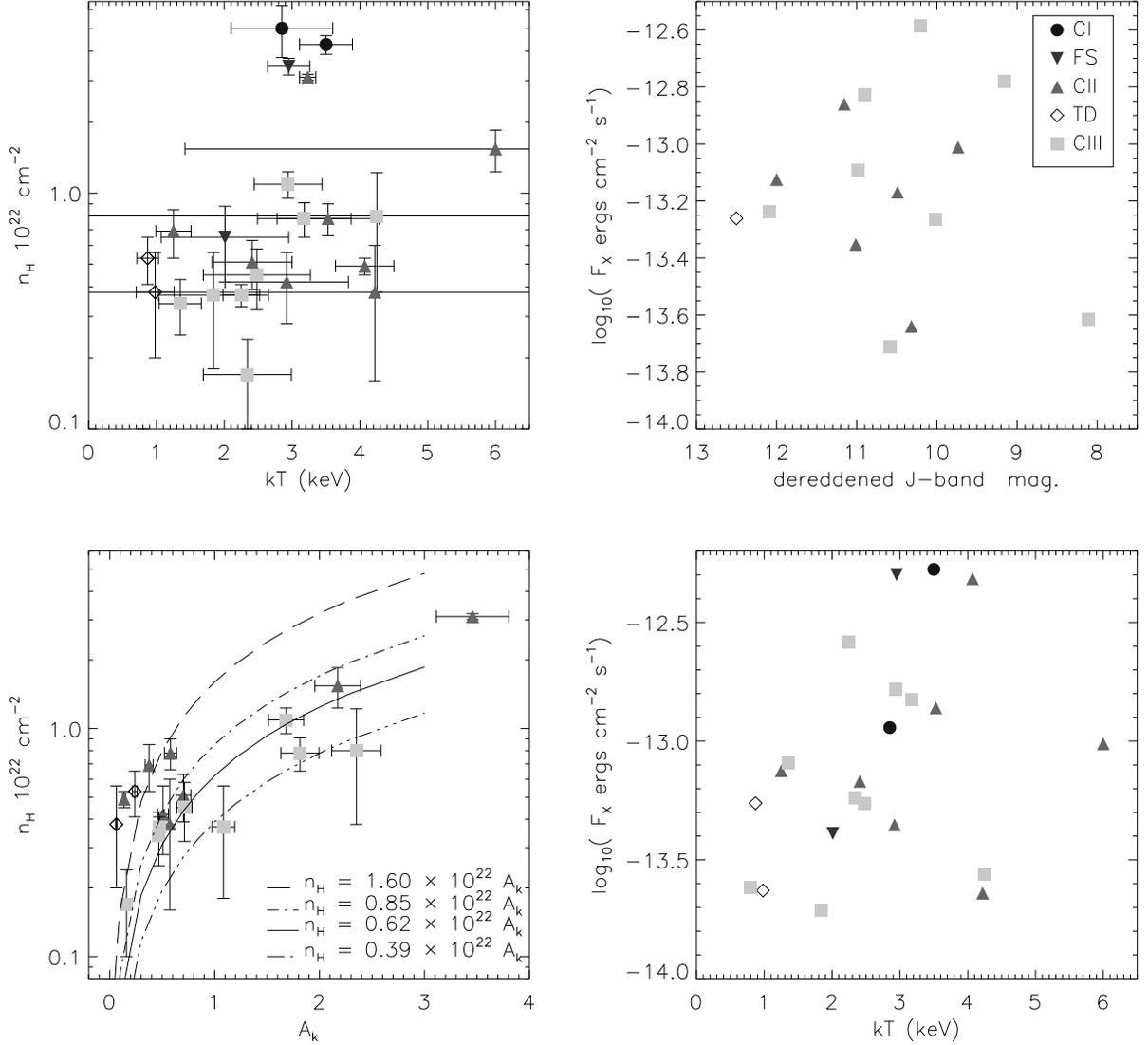}
\caption{ {\it Above left (a):} Plasma temperature against Hydrogen column density. The class 0/I and 
 flat spectrum sources show higher $N_H$, consistent with the presence of an infalling envelope. 
 The class II and III sources do not exhibit a trend in $kT$ with $N_H$. 
 {\it Above right (b):}  X-ray luminosity against dereddened J-band luminosity. A slight trend of increasing 
 flux with increasing luminosity is present. 
 {\it Below left (c):} Hydrogen number density against extinction at K-band. The gas density was calculated 
 from the {\it Chandra} data. The curves indicate the standard $N_H/A_K$ ratio, which does not hold for 
 $A_K > \sim1.5$, and the ratio calculated from these data. 
 {\it Below right (d):} The plasma temperature plotted against the log of the X-ray luminosity. A slight trend 
 of increasing flux with $kT$ is present. 
 The symbols are as follows: class 0/I, circle; flat spectrum, inverted triangle; class II, 
 triangle; transition disk, diamond; class III, squares.   
}
\label{figx}
\end{figure}

Fig.\ref{figx}(a) shows plasma temperature against the column density of Hydrogen in the 
intervening gas between us and the hot plasma.   The class I and flat  spectrum  
sources show column densities five times greater than the class II  and III sources, 
with one exception.  This is not surprising since the class 0/I and  flat spectrum  
sources are surrounded by infalling envelopes.  The one exception is  a class II  
object with an unusually high column density, this may be a source  
where an edge-on disk around the star is occulting the hot plasma.
There does appear to be a correlation of plasma temperature with column density: the
values of these 26  sources show a Spearman rank coefficient of 0.55.  However, the mean values of kT for
$N_{H}$ greater and less than $2\times 10^{22}$~$cm^{-2}$ are $3.13 \pm0.25$~keV  and $2.62 \pm1.34$~keV, respectively,  
indicating that there is no strong dependence.
This indicates that the column density is not biasing our  determination  
of the temperature to higher values by selectively absorbing the soft  X-ray flux.  
Soft X-ray photons are more highly absorbed than hard X-ray photons,  by up to  
90\% when $A_V > 30$ \citep{fei}.  Further more, it shows  that there  
is no significant trend of plasma temperature with evolutionary class.

Fig.\ref{figx}(c) shows the X-ray flux corrected for absorption and $kT$ for  the different 
classes.  Although the class 0/I objects appear slightly more luminous, this may 
be a bias due to the fact that the class 0/I sources are much more highly 
absorbed and  consequently the  weaker class I objects may not have had the 100 
counts we require to perform the analysis of the X-ray spectrum.  
The 8 class II and  9 class III sources  have statistically indistinguishable 
values; the mean values of the class II are  $kT$ $= 3.45 \pm1.31$~keV and 
$F_X$ $= 1.33 \pm1.47 \times 10^{13}$~$erg$~$cm^{-2} s^{-1}$ and the mean value for the class  
III sources is $kT$ $= 2.38 \pm0.96$~keV and $F_X$ $= 0.93 \pm0.77 \times 10^{13}$~$erg$~$cm^{-2} s^{-1}$ .
The mean values for the class I are $kT$ $= 3.17 \pm0.33$~keV and 
$F_X$ $= 3.22 \pm2.08 \times 10^{13}$~$erg$~$cm^{-2} s^{-1}$, for the flat spectrum are $kT$ $= 2.48 \pm0.47$~keV 
and $F_X$ $= 2.72 \pm2.31 \times 10^{13}$~$erg$~$cm^{-2} s^{-1}$, and for the transistion disks are 
$kT$ $= 0.92 \pm0.06$~keV and $F_X$ $= 3.92 \pm1.56 \times 10^{14}$~$erg$~$cm^{-2} s^{-1}$.
Using the combined class II and class III sample  we find only some indication 
of a weak correlation of kT with X-ray flux, with a Spearman rank coefficient  
of 0.39. In comparison, \citet{jef} find a trend of increasing kT with X-ray 
in a combined study of several young clusters.  Finally, the mean  
plasma temperatures for the class 0/I and flat spectrum
sources are not significantly different than those values for class II
and class III sources.  Thus, we find no significant dependence of  
plasma temperature on evolutionary class for the detected X-ray sources.

The X-ray flux of pre-main sequence stars has been found to vary with bolometric luminosity \citep{fei4,cas}.  
Although we cannot measure the bolometric luminosities, we can examine 
the X-ray flux as a function of the dereddened J-band  magnitude, which is 
the  infrared band least contaminated by infrared excess emission and is thus 
a good proxy for photospheric emission  (Fig.\ref{figx}(b)).   
We find again that the class II and class III sources appear indistinguishable.  
In particular,  if we limit our analysis to dereddened $J < 11$ where we detect 90\% of 
the class II sources in X-rays, we find that the $F_X$ $= 6.26 \pm3.06 \times 10^{-14}$~$erg$~$cm^{-2} s^{-1}$ 
for class II and $F_X$ $= 10.76 \pm8.15 \times 10^{-14}$~$erg$~$cm^{-2} s^{-1}$ for class III.
A similar lack of dependence is found in NGC 1333 and $\sigma$ Ori  \citep{get2,her}. 
This suggests that the mechanism generating the X-ray flux is similar during the 
class II and III phases.  
\citet{get2} find a clear relationship between $F_X$ and J-band  magnitude of slope 
$0.47 \pm0.06$ in NGC1333, while \citet{cas} find a slope of 0.30 in the Ophiuchi cloud.  
Our results give a slope of  $0.42 \pm0.17$, which is consistent with these other young 
clusters, although there is a large uncecrtainty due to the scatter in our data.  
In conclusion, we find that there is no clear observational motivation for invoking different 
mechanisms for generating the X-ray emission in the class I, II and III phases.

The combined X-ray and infrared results can be used to calibrate the  relationship 
between gas column density and the extinction measured in the infrared.  We  
calculated $A_K$ for each star using the method of \citet{gut1}, which is based on 
the reddening loci from  \citet{mey} and the extinction law of \citet{fla}.  
These values can  be compared to the column density of hydrogen atoms, $N_H$,  
which is calculated from the  inferred  absorption of the X-ray emission. Previous 
measurements of  this value lead  to an approximately linear fit of 
$1.6\times10^{21} A_{V}$ \citep{vuo,fei} for stellar sources, while the value for the 
diffuse ISM ranges from $1.8-2.2 \times10^{21} A_{V}$ \citep{vuo,gore}.  
Fig.\ref{figx}(c) plots the hydrogen  
column density against the calculated exctinction in the K-band.  As shown in this 
plot,  the \citet{vuo} relationship agrees with our data for $A_K < 1$,  
but diverges from the points with $A_K > 1$. A similar divergence has also been found 
in the more deeply embedded stars in the RCW 108 region \citep{wol2}.

Interestingly, the class II sources always show a higher value of $N_H$ for 
a given $A_K$. The colors of the class II objects can be affected by both absorption 
and scattering of the light from the central source by a flared disk; the extent of 
this effect  would be determined, predominantly, by the inclination of the disk and 
the sizes of  grains in the disk \citep{whi2}. For example, the colors of an edge-on 
disk system,  where the central star is highly obscured by the disk,  would be 
dominated by scattered light.    In such a case, the disk absorbs both the infrared 
and the X-ray emission  from the star, but the scattered infrared light reduces the 
observed reddening of the star and hence the inferred  value of $A_K$.
To eliminate the uncertainties in the determination of $A_K$ due to circumstellar 
material, we fit only the class III objects. The resulting relationship is 
$N_H = 0.63\pm0.23 \times 10^{22} A_K$.   Assuming a standard conversion 
of $A_K = 0.1 A_V$ \citep{rie2}, this slope is 2.5 times lower than that derived 
by previous authors. In comparison, a fit to the class II objects results in 
$N_H = 1.3\pm0.93 \times 10^{22} A_K$.  The poor fit and higher slope are further indication that 
the disks around these objects affect the measurement of $A_K$, possibly leading 
to spuriously low values of $A_K$.
We now focus on the slope derived from the class III objects, the possible systematic 
uncertainties in that slope, and a possible explanation for the low value of the slope.

{\it Systematic underestimation of $N_H$:} 
If the measurements of $N_H$ were systematically underestimated by a factor of 2.5, then 
this would lead to the shallower slope observed.  
However, the determinations of the $N_H$ values appear to be robust.  
We calculated $N_H$ only for those sources with relatively high S/N ($> 100$ counts).  These values 
of $N_H$ were found to be consistent using two different models for the plasma.  
A bias could occur in the determination of $N_H$  when the soft X-ray emission is totally 
absorbed; however inspection  shows that this is not the case here.
Furthermore, values of $N_H$ calculated by our software for stars in the Orion COUP survey are 
within 5\% of the values published in \citet{get}; this is further validation of the 
accuracy of our analysis.

{\it Sytematic overestimation of $A_K$:}
The value of $A_K$ is determined by the selective absorption in the near-infrared bands 
\citep{gut1}; this selective absorption was converted to $A_K$ by using an assumed value 
of $A_{K}/E(H-K)$.  We use the \citet{ind} reddening law where the adopted ratio of 
$A_{H}/A_{K} = 1.55$. This is  equivalent to a near-IR general to selective absorption of 
$A_{K}/E(H-K) = 1.82$.  Changes in the reddening law in the molecular cloud could change 
this ratio.  In molecular clouds, \citet{fla} report a flattening in the infrared reddening law for 
$\lambda > 3.6$~${\mu}m$ compared to the diffuse 
ISM reddening law measured by \citet{ind}; however, they found no significant difference 
in the reddening law for $\lambda = 1 - 2.1$~${\mu}m$ between the molecular clouds and the diffuse ISM.
Alternatively, \citet{nis} report $A_{K}/E(H-K)$ of 1.44, lower than the value in \citet{ind}, 
and they find evidence of variations in this value toward different lines of sight toward the galactic center. 
If the slope found in our data were wholly due to an overestimation of $A_K$, then the  value of 
$A_{K}/E(H-K)$ would have to decrease to 0.73, and the value of $A_{H}/A_{K}$ must change to 2.37. 
This value of $A_{H}/A_{K}$ is higher than the values in the literature and would require a steepening 
of the reddening law between 1 \& 2~${\mu}m$.

{\it Growth and Coagulation of Dust Grains:} The ratio $N_H/A_K$ can be affected by increases 
in the sizes of dust grains; this would simply increase the value of $A_K$ and hence 
reduce the ratio of $N_H/A_K$.  The value of $A_K$ is given by:
\begin{eqnarray*}
A_K = 1.086 N_d \pi \int a^2 Q_{ext} n(a) da
\end{eqnarray*}
where $N_d$ is the column density of dust, $a$ is the radius of a dust grain, and $Q_{ext}$ 
is the extinction efficiency \citep{whit}. In the limit of $\lambda < a$, $Q_{ext}$, which 
would be the case for a standard grain size distribution \citep{mat} of
$n(a) \propto 0.005 < a_d^{-3.5} < 0.25 {\mu}m$,
\begin{eqnarray*}
Q_{ext} = \frac{8 \pi a}{\lambda} Im\left\{\frac{m^2 - 1}{m^2 + 2} 
\right\} +
\frac{8}{3}(\frac{2 \pi a}{\lambda})^4 \left\vert\frac{m^2 - 1}{m^2 +
2}\right\vert^2
\end{eqnarray*}
where $m$ is the refractive index.  Thus, there is a strong dependence on the size of 
the grain, and the value of $A_K$ can be changed simply by growing grains to larger sizes; 
for example the scattering term in $Q_{ext}$ goes as $(\frac{2 \pi a}{\lambda})^4$, hence a 25\%
increase in grain size can increase the scattering efficiency by a factor of 2.5.

Grain growth in molecular clouds may occur through both coagulation and the accretion of grain 
mantles; these processes would be accelerated in the dense molecular gas found in the Serpens 
cloud core.  For example, \citet{jura} explained  changes in the $N_H/A_V$ ratio in Ophiuchus
through the coagulation of grains. The collision time between grains is \citep{jura}
\begin{eqnarray*}
t_{col} = (10^{-16} n_{H} [cm^{-3}] v_d [km~s^{-1}])^{-1}
\end{eqnarray*}
For a $ n_{H} = 10^5 cm^{-3}$ and a velocity of 0.1~km~s$^{-1}$ \citep{jura}, the collision 
timescale is 30,000 years.  Thus it is likely that a dust grain in the Serpens cloud would 
have had multiple collisions with other grains. Alternatively, the growth can be through the 
accretion of mantles of volatiles such as H$_2$O,  NH$_3$, O$_2$, N$_2$ and CO.  The accretion 
timescale is given by
\begin{eqnarray*}
t_m = \frac{2.5 s \Delta a}{S n (k T_g m)^\frac{1}{2}}
\end{eqnarray*}
where s is the specific density of the mantle, $\Delta a$ is the change in the size of the dust 
grain due to the mantle, $S$ is the sticking coefficient ($\sim 1$), $T_g$ is the temperature 
of the gas, $n$ is the density of the molecule forming the mantle, and $m$ is the mass of the 
molecule forming the mantle \citep{whit}.  Thus the rate of mantle accretion is proportional to 
the density of the gas; again, the high density of the gas in the Serpens cloud would accelerate 
the rate of mantle accretion. A detailed calculation of the rate requires consideration of the 
gas phase abundances of the different possible volatiles, and is beyond the scope of this paper.

Self consistent modeling  of the extinction over multiple wavelength bands is needed 
to study the impact of grain growth on both the relationship between $N_H$ and $A_K$ 
and the shape of extinction  law with which $A_K$ is derived.  Further observational 
work is also needed to confirm our observed $N_H/A_K$ slope using a larger number of 
class III objects from multiple star forming regions. However, this study demonstrates 
the value of using combined Chandra, near-IR, and Spitzer observations to identify  
deeply embedded class III objects and search for variations in the extinction law  
within molecular clouds.

\section{\bf Conclusion}

We have undertaken an intensive study of the Serpens Cloud Core in the mid-IR with {\it Spitzer} 
and in the near-IR with 2MASS. We have merged these data with {\it Chandra} X-ray observations and 
investigated various X-ray properties of protostars and pre-main sequence stars. 
These data provide perhaps the most complete census to date of the young stellar objects in one of 
the nearest embedded clusters to the Sun.

\begin{itemize}
\item We identify 138 young stellar objects, after removing contamination.
These include 38 protostars (22 class 0/I, 16 Flat Spectrum) and 100 pre-main 
sequence stars (62 class II, 17 transition disks, and 21 class III).
The ratio of protostellar objects (class 0/I and flat spectrum) to pre-main 
sequence stars (class II, transition disk, and class III) is 38:100 
or 29\% of the total number of identified members. The ratio of protostars to 
pre-main sequence stars with disks is 48\%. These high numbers indicate that 
Serpens is unusually rich in protostars. We estimate that as many as 44 class 
III sources may be missing, implying a total membership of 182.

\item We have searched for {\it Spitzer} counterparts to 14 (sub)mm sources in 
the Serpens cluster.  Seven are detected as point sources at 3.6-8.0~${\mu}m$, 
a total of 10 have been detected at 24~${\mu}m$, and an additional two show 
nebulosities at $\lambda \le 24$~${\mu}m$. All but one are coincident with 
70~${\mu}m$ emission, although for many of the protostars, the low angular 
resolution of the 70~${\mu}m$ data precludes accurate photometry of the sources.

\item We find that the spatial distribution of the protostars is distinctly  
different than the pre-main sequence stars. While the class 0/I sources and 
most of the flat spectrum sources are concentrated in the central molecular 
gas filament, the class II, transition disk and class III sources are  
distributed over a much more extended region. The median projected spacing 
of the Class 0/I sources is only 5000~AU; these sources are highly clustered.
In contrast, the distribution of the remaining classes are statistically 
indistinguishable from a random distribution.

\item The protostars are concentrated in two groups coincident with the two  
dominant molecular gas clumps, as has been previously reported. The density 
of protostars in these groups suggest that the volumes of gas from which 
these protostars accrete may touch or overlap.  The spacing is also similar 
to both the Jean's length and the radii of critical Bonner Ebert spheres. 
Based on published ambient gas densities and velocities of the protostars 
in the northern group, we find  that competitive accretion could be occurring 
in this groups. 
Furthermore, collisions between protostellar envelopes may be common.

\item Of the 138 sources, 60 exhibit detectable X-ray emission.  Nine of the 
detections are associated with class I sources (43\% of members in class), 
a further eight with flat spectrum members (56\%). Twenty-one detections have 
class II counterparts (45\%), while only two of the transition disks are 
detected in X-rays (25\%).  Twenty class III sources were identified solely by their 
X-ray emission. There appears to be no evidence of a trend in the detection 
rate of sources by evolutionary class. The fraction of X-ray detected pre-main 
sequence stars (i.e. class II, transition disks and class III sources) with disks is 51\%.

\item We find no evidence for a dependence of the temperature of the X-ray 
emitting plasma on the evolutionary class, and find no difference in the X-ray 
luminosity of detected class II and classs III objects. 
Consequently, there is no evidence that the generation mechanism 
of the X-ray emission depends on the presence of an infalling envelope or 
circumstellar disk.

\item We find an observed relationship of $N_H = 0.63\pm0.23 \times 10^{22} A_K$, 
which is much shallower than found in other studies.  
The most likely explanation for this is a change in the near-IR reddening law 
that would affect the shape and magnitude of the reddening law. Such changes 
could result from grain coagulation and/or grain growth from the accretion of 
mantles of volatiles; both such processes may occur relatively quickly due to 
the high gas (and presumably dust grain) densities in the Serpens cloud core.

\end{itemize}

This work is based on observations made with the {\it Spitzer} Space Telescope (PID 6, PID 174), 
which is operated by the Jet Propulsion Laboratory, California Institute of Technology under NASA 
contract 1407. Support for this work was provided by NASA through contract 1256790 issued by 
JPL/Caltech. Support for the IRAC instrument was provided by NASA through contract 960541 issued 
by JPL.
This publication makes use of data products from the Two Micron All Sky Survey, which is a 
joint project of the University of Massachusetts and the Infrared Processing and Analysis 
Center/California Institute of Technology, funded by the National Aeronautics and Space 
Administration and the National Science Foundation.
This research has made use of the NASA/IPAC Infrared Science Archive, which is operated by 
the Jet Propulsion Laboratory, California Institute of Technology, under contract with the 
National Aeronautics and Space Administration.
E. Winston would like to thank the Irish Research Council for Science, Engineering, 
and Technology (IRCSET) for funding and support.

\clearpage

\begin{deluxetable}{ccccc}
\tabletypesize{\scriptsize}
\tablecaption{Table of Source Identification Details}
\label{sid}
\tablewidth{0pt}
\tablehead{\colhead{}  &  \colhead{IR-excess}  & \colhead{X-ray Detected}   & \colhead{Extremely Red} & \colhead{Total} }
\startdata
Identified      & 200   &  67  &  3  &  229 \\
Contam.         &  87   &   7  &  1  &   91 \\
YSOs            & 113   &  60  &  2  &  138 \\
\enddata
\end{deluxetable}

\clearpage

\begin{deluxetable}{cccc}
\tabletypesize{\scriptsize}
\tablecaption{Table of Source Classifications}
\label{tbl2}
\tablewidth{0pt}
\tablehead{\colhead{Class} & \colhead{IR-field}  & \colhead{IRX-field}  & \colhead{X-ray Detections} }
\startdata
0/I     &  22  &  22  &   9 ( 43\%)  \\
Flat    &  16  &  15  &   8 ( 53\%)  \\
II      &  62  &  44  &  20 ( 46\%)  \\
Trans.  &  17  &   8  &   2 ( 25\%)  \\
III     &  21  &  21  &  21   \\
\enddata
\end{deluxetable}

\clearpage

\begin{deluxetable}{cccccc}
\tabletypesize{\scriptsize}
\tablecaption{Table of class distribution comparison: Kolmogorov-Smirnov probabilities.}
\label{tprob}
\tablewidth{0pt}
\tablehead{
\colhead{}   & \colhead{I}  & \colhead{Flat} & \colhead{II}   & \colhead{Trans.}  & \colhead{III} }
\startdata
Random     &  $2\times10^{-7}$    &  0.05                &  0.13                &  0.67                & 0.21                \\
I          &  \nodata             &  $5.7\times10^{-4}$  &  $1.2\times10^{-5}$  &  $4.6\times10^{-5}$  & $3.0\times10^{-5}$  \\
Flat       &  $5.7\times10^{-4}$  &  \nodata             &  0.45                &  0.28                & 0.73                \\
II         &  $1.2\times10^{-5}$  &  0.45                &  \nodata             &  0.016               & 0.60                \\
Trans.     &  $4.6\times10^{-5}$  &  0.28                &  0.016               &  \nodata             & 0.35                \\
\enddata
\end{deluxetable}

\clearpage

\center
\begin{deluxetable}{cccccccc}
\tablecolumns{8}
\tabletypesize{\scriptsize}
\tablewidth{0pt}
\tablecaption{Table of Coordinates \& Other Identifiers \protect\label{tableids}}
\tablehead{\colhead{{\it Spitzer} ID}  & \colhead{RA (2000)}   & \colhead{Dec (2000)}  & \colhead{EC\,\tablenotemark{a}}  & \colhead{Kaas\,\tablenotemark{b}} & \colhead{Class 0\,\tablenotemark{c}} & \colhead{{\it Chandra} ID} & \colhead{GFG ID\,\tablenotemark{d}}}
\startdata
1  &  18:30:02.09  &  1:13:58.81  &  CK7  &  \nodata  &  \nodata  &  \nodata  &  \nodata  \\ 
2  &  18:29:57.83  &  1:12:51.62  &  EC95  &  Kaas317/345  &  \nodata  &  72  &  53  \\ 
3  &  18:29:59.57  &  1:11:58.80  &  \nodata  &  Kaas330  &  \nodata  &  25  &  60  \\ 
4  &  18:29:58.76  &  1:14:26.47  &  EC103  &  Kaas326  &  \nodata  &  23  &  \nodata  \\ 
5  &  18:29:47.01  &  1:16:26.82  &  EC27  &  Kaas237  &  \nodata  &  \nodata  &  \nodata  \\ 
6  &  18:29:49.24  &  1:16:31.74  &  \nodata  &  Kaas249  &  \nodata  &  \nodata  &  \nodata  \\ 
7  &  18:29:49.58  &  1:17:05.92  &  EC76  &  Kaas254  &  S68Nb  &  17  &  28  \\ 
8  &  18:29:57.59  &  1:13:00.44  &  SVS4-10  &  Kaas312  &  \nodata  &  12  &  46  \\
\enddata

\tablenotetext{a}{Identifiers from \citet{eir1}.}
\tablenotetext{b}{Identifiers from \citet{kaa}.}
\tablenotetext{c}{Identifiers from \citet{tes1,wil1,dav1}.}
\tablenotetext{d}{Identifiers from \citet{gio}.}

\end{deluxetable}

\clearpage

\begin{deluxetable}{ccccccccccccc}
\tablecolumns{13}
\tabletypesize{\scriptsize}
\rotate
\tablewidth{0pt}
\tablecaption{Table of Young Stellar Objects in Serpens \protect\label{tablephot}}
\tablehead{
\colhead{ID} & \colhead{J}  & \colhead{H} 
 & \colhead{K} & \colhead{3.6${\mu}m$} & \colhead{4.5${\mu}m$}  & \colhead{5.8${\mu}m$}  & \colhead{8.0${\mu}m$}  & \colhead{24.0${\mu}m$}  & \colhead{Class} & \colhead{CID\,\tablenotemark{a}} & \colhead{$\alpha_{IRAC}$\,\tablenotemark{b}} 
 & \colhead{A$_{K}$\,\tablenotemark{c}}  }

\startdata
1  &  $>$16.044  &  14.921$\pm$0.118  &  13.389$\pm$0.046  &  10.651$\pm$0.003  &  9.806$\pm$0.002  &  9.081$\pm$0.003  &  8.207$\pm$0.002  &  4.030$\pm$0.004   &  0/1  &  \nodata  &  -0.06  &   \nodata \\
2  &  15.752$\pm$0.093  &  12.455$\pm$0.045  &  10.518$\pm$0.029  &  8.646$\pm$0.004  &  7.680$\pm$0.003  &  6.874$\pm$0.004  &  5.639$\pm$0.004  &   1.099$\pm$0.013  &  0/1  &  72  &  0.56  &    \nodata \\
3  &  $>$17.913  &  $>$17.371  &  14.649$\pm$0.099  &  8.959$\pm$0.003  &  7.263$\pm$0.002  &  6.195$\pm$0.002  &  5.309$\pm$0.001  &  1.504$\pm$0.006  &  0/1   &  25  &  1.27  &  \nodata  \\
4  &  $>$15.185  &  14.025$\pm$0.069  &  11.843$\pm$0.042  &  9.485$\pm$0.004  &  8.383$\pm$0.005  &  7.332$\pm$0.005  &  6.619$\pm$0.005  &  2.471$\pm$0.021   &  0/1  &  23  &  0.46  &   \nodata  \\
5  &  $>$17.731  &  $>$15.836  &  13.403$\pm$0.041  &  11.041$\pm$0.004  &  10.218$\pm$0.005  &  9.614$\pm$0.007  &  8.910$\pm$0.004  &  5.463$\pm$0.022  &  0/1  &   \nodata  &  -0.44  &  \nodata  \\
6  &  $>$18.248  &  $>$17.326  &  13.446$\pm$0.04  &  10.742$\pm$0.004  &  9.322$\pm$0.004  &  8.255$\pm$0.003  &  7.051$\pm$0.006  & 1.665$\pm$0.022   &  0/1  &   \nodata  &  1.35  &  \nodata  \\
7  &  $>$16.889  &  15.658$\pm$0.145  &  12.67$\pm$0.036  &  9.621$\pm$0.002  &  8.282$\pm$0.003  &  7.302$\pm$0.002  &  6.447$\pm$0.002  &  2.741$\pm$0.009  &   0/1  &  17  &  0.76  &   \nodata  \\
8  &  $>$16.454  &  $>$14.104  &  13.055$\pm$0.085  &  9.188$\pm$0.007  &  7.665$\pm$0.004  &  6.747$\pm$0.004  &  5.977$\pm$0.007  &  1.431$\pm$0.014  &  0/1   &  12  &  0.77  &  \nodata  \\
\enddata

\tablenotetext{a}{{\it Chandra} X-ray Idenifier number.}
\tablenotetext{b}{Slope of SED calculated from available IRAC data points. Note slopes of FS objects fall between -0.3$> \alpha >$0.3 when 24${\mu}m$ point is included.}
\tablenotetext{c}{This source was saturated at 24~${\mu}m$, the photometry was fit using the PSF wings as described in text.}
\tablenotetext{d}{Photometry at this wavelength is saturated.}
\tablenotetext{e}{Slope \& classification uncertain due to lack of photometry in all {\it Spitzer} bands. The IR-excess may be due 
to a mismatch between  the 2MASS \& IRAC photometry.}

\end{deluxetable}

\clearpage

\begin{center}
\begin{deluxetable}{cccccccccccccccc}
\tablecolumns{16}
\tabletypesize{\scriptsize}
\rotate
\tablewidth{0pt}
\tablecaption{Table of submm and mm sources \protect\label{tablesmm}}
\tablehead{
\colhead{Identifer}  & \colhead{RA\,\tablenotemark{a}} & \colhead{Dec\,\tablenotemark{a}}  & \colhead{{\it Spit.} RA}  & 
\colhead{{\it Spit.} Dec} & \colhead{3.6${\mu}m$} & \colhead{4.5${\mu}m$} & \colhead{5.8${\mu}m$} & \colhead{8.0${\mu}m$} & 
\colhead{24${\mu}m$} & \colhead{70${\mu}m$}   & \colhead{450${\mu}m$\,\tablenotemark{b}} & \colhead{850${\mu}m$\,\tablenotemark{b}}   & \colhead{3$mm$\,\tablenotemark{c}} & \colhead{3.5$cm$\,\tablenotemark{d}} &
\colhead{CXO\,\tablenotemark{e}} \\
\colhead{}  & \colhead{J2000} & \colhead{J2000}  & \colhead{J2000}  & \colhead{J2000} &
\colhead{(mJy)} & \colhead{(mJy)} & \colhead{(mJy)} & \colhead{(mJy)} &  \colhead{(mJy)} & \colhead{(mJy)} & 
\colhead{(mJy)} & \colhead{(mJy)} & \colhead{(mJy)}     & \colhead{(mJy)}  & \colhead{I.D.} 
}
\startdata
SMM1 & 18:29:49.69 & 1:15:20.55  & 18:29:49.66 & 1:15:21.56 & \nodata & \nodata & \nodata & 10.7\,\tablenotemark{f} & 1538 & 28034 & 35700 & 6100  & 233. & 7.54  & \nodata \\
SMM2 &	18:30:00.33  & 1:12:55.33  &  \nodata  & \nodata   & \nodata &	\nodata & \nodata & \nodata & \nodata &	\nodata & 3400 & 700 &   \nodata & \nodata & \nodata \\
SMM3 &	18:29:59.41  &  1:14:01.25 &  18:29:59.27 & 1:14:02.90 & \nodata & \nodata & \nodata &	\nodata & 472 & \nodata & 7100  & 1500  &  \nodata & \nodata & \nodata \\
SMM4 &	18:29:56.73  &   1:13:17.07  &  18:29:56.69  & 1:13:20.71 & 0.29\,\tablenotemark{f} & 2.53\,\tablenotemark{f} &	2.25\,\tablenotemark{f} & \nodata & \nodata  & \nodata & 10800  & 2900  &  \nodata & 0.12 & \nodata \\
SMM5 &	18:29:51.16  &  1:16:40.65  &  18:29:51.14 & 1:16:40.47 & 27.8 & 59.4 &	125 &	183   &  1376 & 3705  & 3900  &  900  & 20.7 & 0.15   &  18 \\
SMM6 &	18:29:57.71  &  1:14:08.12  &  18:29:57.73 & 1:14:06.03 & 2269 & 3163 &	4899 &	5897   & 9676  & 5452  &  1900  & 500  &   \nodata  & 0.17 & 70  \\
SMM8 &	18:30:01.79  &  1:15:09.43  &  18:30:01.53 & 1:15:14.90 & \nodata & \nodata & \nodata &	1.55\,\tablenotemark{g}   &	21.6 &	\nodata & 1300  & 400  &  \nodata  & \nodata & \nodata  \\
SMM9 & 18:29:48.16  &   1:16:44.43  &  18:29:48.10 & 1:16:44.97  & 1.82 & 6.41 & 11.1 &	15.1    & 289 & 5664 & 9100  & 1600  &  36.1 & 0.28 & \nodata  \\
\enddata
\tablenotetext{a}{Coordinates from \citet{tes1}, except S68Nb-d taken from \citet{wil2}. Brackets indicate possible offset location of SMM11. }
\tablenotetext{b}{Peak Fluxes in mJy/beam from \citet{dav1}, assuming all flux from source falls in one beam.}
\tablenotetext{c}{Coordinates \& Peak Fluxes from \citet{wil2}.}
\tablenotetext{d}{Peak Fluxes from \citet{eir2}.}
\tablenotetext{e}{The {\it Chandra} Identifier number, as in Table~\ref{tableids}.}
\tablenotetext{f}{Extended nebulous source.}
\tablenotetext{g}{{\it Spitzer} fluxes associated with sources contaminated by flux from nearby bright objects.}
\tablenotetext{h}{Detected in \citet{tes1}, no flux listed.}

\end{deluxetable}
\end{center}

\clearpage

\clearpage

\clearpage

\clearpage

\clearpage

\clearpage

\clearpage

\clearpage

\clearpage

\clearpage

\clearpage

\clearpage

\clearpage

\clearpage

\clearpage

\clearpage

\end{document}